\documentclass[epj]{svjour}
\usepackage{graphics}
\newcommand{\sitei}{\vec{i}}
\newcommand{\sitej}{\vec{j}}

\newcommand{\sitek}{\vec{k}}
\newcommand{\siteq}{\vec{q}}

\newcommand{\sitezero}{\vec{0}}
\newcommand{\I}{{\rm i}}
\newcommand{\diff}{{\rm d}}
\newcommand{\sgn}{{\rm sgn}}

\begin{document}
\title{Fourth-Order 
Perturbation Theory for the Half-Filled Hubbard Model in Infinite Dimensions}
\author{Florian Gebhard \inst{1}
\and 
Eric Jeckelmann \inst{2}
\and
Sandra Mahlert \inst{1}
\and
Satoshi Nishimoto \inst{1}
\and
Reinhard M.~Noack \inst{3}
}                     
%
%
\institute{Fachbereich Physik, Philipps-Universit\"at Marburg,
D-35032 Marburg, Germany
\and 
Institut f\"ur Physik, Johannes-Gutenberg Universit\"at Mainz,
D-55099 Mainz, Germany
\and
Institut f\"ur Theoretische Physik~III, Universit\"at Stuttgart,
D-70550 Stuttgart, Germany
}
\date{Received: \today / Revised version: }
%
\abstract{We calculate the zero-temperature self-energy
to fourth-order perturbation theory in the Hubbard interaction~$U$
for the half-filled Hubbard model in infinite dimensions.
For the Bethe lattice with bare bandwidth $W$,
we compare our perturbative results for the self-energy,
the single-particle density of states, and the momentum
distribution to those from 
approximate analytical and numerical studies of the model.
Results for the density of states from perturbation theory 
at $U/W=0.4$ agree very well with those from 
the Dynamical Mean-Field Theory treated with
the Fixed-Energy Exact Diagonalization and 
with the Dynamical Density-Matrix Renormalization Group. 
In contrast, our results reveal the limited resolution of 
the Numerical Renormalization Group approach in treating
the Hubbard bands.
The momentum distributions from all approximate studies of the model
are very similar in the regime where perturbation theory
is applicable, $U/W \le 0.6$. Iterated Perturbation Theory
overestimates the quasiparticle weight above 
such moderate interaction strengths.
} 
\maketitle
\section{Introduction}
\label{intro}

The Hubbard model serves as a paradigm for strongly correlated
electron systems because it combines the two essential
aspects of electrons in solids in a simplistic way.
It describes spin-1/2 electrons moving on a lattice
(bandwidth~$W$) which interact only locally with strength~$U$ 
(Hubbard interaction).
Despite its structural simplicity
it is thought to encompass a rich phase diagram.

At half band-filling, when there is on average one electron 
per lattice site, the Hubbard model contains a zero-temperature 
phase transition from a metal to an 
insulator, irrespective of any symmetry 
breaking~\cite{Mottbook,Gebhardbook,RMP,Imada}. 
In the limit of large lattice dimensions~\cite{MV},
the precise nature of this Mott-Hubbard transition
is not entirely clear. An exact solution
of the problem is not possible, and various approximate treatments
have led to two conflicting scenarios;
for a review,
see Refs.~\cite{Gebhardbook,RMP}. For more recent treatments, 
see Refs.~\cite{RDA,7Schwaben,7Schwabenreply,Krauth,%
BullaPRL,BullaVolli,BullaLDMFT,EastwoodGebhard}.

\begin{description}
\item[\emph{Discontinuous Transition.}] \mbox{} \hfill

The gap jumps to a finite value
when the density of states at the Fermi energy becomes zero at some critical
interaction strength~$U_{{\rm c},2}$; the gap is preformed
above $U_{{\rm c},1}<U_{{\rm c},2}$, and the
co-existing insulating state is higher in energy 
than the metal.

\item[\emph{Continuous Transition.}] \mbox{} \hfill
\nopagebreak[4]

The gap opens continuously
when the density of states at the Fermi energy becomes zero,
$U_{{\rm c},1}=U_{{\rm c},2}\equiv U_{\rm c}$.
\end{description}

This situation calls for the development and application
of systematic and controlled techniques such as 
high-order perturbation theory
in strong and weak coupling. Recently, some of
us~\cite{EastwoodGebhard} 
investigated the Mott-Hubbard insulator
on a Bethe lattice in the limit of large coordination number
analytically and numerically.
In the present work, after the introduction of some definitions 
in Sect.~\ref{sec:Def},
we calculate the one-particle Green function of the metallic state
at half band-filling
up to and including ${\cal O}[(U/W)^4]$ in Sect.~\ref{sec:PTU}; 
corrections are of the order $(U/W)^6$ due to particle-hole symmetry.
Perturbation theory to fourth order 
is found to converge very well at $U=0.4 W$,
but it begins to fail at $U\approx 0.64 W$.
In Sect.~\ref{sec:DMFT}, we compare our perturbative results
for the density of states 
with those of the Dynamical Mean-Field Theory (DMFT),
analyzed within two recently developed numerical schemes,
the Fixed-Energy Exact Diagonalization 
(FE-ED)~\cite{EastwoodGebhard}
and the Dynamical Density-Matrix Renormalization Group
(DDMRG)~\cite{Jeckelmann,JeckelNishimoto}. 
The DMFT, which becomes exact in the limit of infinite dimensions,
requires the self-consistent solution of a single-impurity
Anderson model on $n_s\to\infty$ bath sites. We employ the
FE-ED for $n_s\leq 15$ and the
DDMRG on up to $n_s=64$ sites. We find very good agreement with our
perturbative results for $U=0.4 W$, and can attribute deviations
at $U=0.6 W$ to the limited accuracy of fourth-order perturbation theory.
Therefore, we conclude that the DDMRG provides a reliable description
of the correlated metal at all frequencies.
The comparison is less favorable with results from 
the Numerical Renormalization Group (NRG)~\cite{BullaPRL} 
and Iterated Perturbation Theory (IPT) which show noticeable
deviations for intermediate energies, e.g., in the formation
of the Hubbard bands.

In Sect.~\ref{sec:RDA}, we compare our results
for the momentum distribution $n(\epsilon)$
and the quasi-particle weight $Z(U)$ with those of the
Random Dispersion Approximation (RDA), the NRG and IPT.
Within the region of validity of our perturbation expansion, $U\leq 0.6$,
the results are almost identical except for the IPT which 
already overestimates 
the quasi-particle weight at weak to moderate interaction strengths.
Conclusions, Sect.~\ref{sec:conclusions}, and two appendices close
our presentation.

\section{Definitions and basic properties}
\label{sec:Def}

In this section, we discuss the basic properties of 
lattice electrons in the limit of infinite dimensions.
We define the Hubbard Hamiltonian, the one-particle Green function,
and some related one-particle quantities.

\subsection{Hamilton operator}
\label{subsec:Hamilt}

We investigate spin-1/2 electrons on a lattice. Their
motion is described by
\begin{equation}
\hat{T} =
\sum_{\sitei,\sitej;\sigma} t_{\sitei,\sitej} 
\hat{c}_{\sitei,\sigma}^+\hat{c}_{\sitej,\sigma} \; ,
\end{equation}
where $\hat{c}^+_{\sitei,\sigma}$,
$\hat{c}_{\sitei,\sigma}$ are creation and annihilation operators for
electrons with spin~$\sigma=\uparrow,\downarrow$ on site~$\sitei$.
Here $t_{\sitei,\sitej}$ are the electron transfer amplitudes between
sites $\sitei$~and $\sitej$, and $t_{\sitei,\sitei}=0$.
Since we are ultimately interested in the Mott-Hubbard transition
we consider a half-filled band
where the number of electrons~$N$ equals the
number of lattice sites~$L$ exclusively.

For lattices with translational symmetry, we have $t_{\sitei,\sitej}=
t(\sitei-\sitej)$, and 
the operator for the kinetic energy is diagonal in momentum space,
\begin{eqnarray}
\hat{T} &=& \sum_{\sitek;\sigma} \epsilon(\sitek) 
\hat{c}^+_{\sitek,\sigma}\hat{c}_{\sitek,\sigma}
\nonumber \; ,
\\
\epsilon(\sitek) &=& \frac{1}{L} \sum_{\sitei,\sitej} t(\sitei-\sitej)
e^{-\I (\sitei-\sitej) \sitek} \; . 
\label{disp}
\end{eqnarray}
The density of states for non-interacting electrons is then given by
\begin{equation}
\rho(\epsilon)= \frac{1}{L} \sum_{\sitek} \delta(\epsilon-\epsilon(\sitek))
 \nonumber \; .
\end{equation}
The $m$-th moment of the density of states is defined by
\begin{equation}
\overline{\epsilon^m} = \int_{-\infty}^{\infty} 
\diff\epsilon \, \epsilon^m \rho(\epsilon) \;,
\label{energymoments}
\end{equation}
and $\overline{\epsilon}=t(\sitezero)=0$.

In the limit of high lattice dimensions and for translationally
invariant systems, the Hubbard model is characterized by $\rho(\epsilon)$
alone, i.e.,
higher-order correlation functions 
in momentum space factorize, e.g.,~\cite{PvDetal}
\begin{eqnarray}
\rho_{\siteq_1,\siteq_2}(\epsilon_1,\epsilon_2)
&\equiv&
\frac{1}{L} \sum_{\sitek} \delta(\epsilon_1-\epsilon(\sitek+\siteq_1))
\delta(\epsilon_2-\epsilon(\sitek+\siteq_2))
\nonumber 
\\
&=&
\rho(\epsilon_1) [ \delta_{\siteq_1,\siteq_2} 
\delta(\epsilon_1-\epsilon_2)
%
+ (1-\delta_{\siteq_1,\siteq_2}) \rho(\epsilon_2)] \, .
\nonumber \\
&&\label{RDADq} 
\end{eqnarray}
This observation is the basis for the Random Dispersion Approximation
(RDA) which becomes exact in
infinite dimensions for paramagnetic systems where
nesting is ignored~\cite{Gebhardbook,RDA}; see Sect.~\ref{sec:RDA}.

In the following, we assume a symmetric bare density of states,
$\rho(-\epsilon)=\rho(\epsilon)$, of width~$W$.
For our explicit calculations we shall later use 
the semi-circular density of states
\begin{equation}
\rho_0(\omega)= \frac{2}{\pi W}\sqrt{4 -\left(\frac{4\omega}{W}\right)^2\,}
\quad , \quad   (|\omega|\leq W/2) \; ,
\label{rhozero}
\end{equation}
where $W\equiv 4t$ is the bandwidth. 
In the following, we shall set $t\equiv 1$
as our energy unit if not otherwise explicitly stated.
This density of states is realized for non-interacting
tight-binding electrons on a Bethe lattice of connectivity 
$Z\to\infty$~\cite{Economou},
i.e., each site is connected to $Z$~neighbors without generating closed loops,
and the electron transfer is restricted to nearest-neighbors,
$t_{\sitei,\sitej}=-1/\sqrt{Z}$ for $\sitei$~and $\sitej$~being 
nearest neighbors and
zero otherwise. The limit $Z\to\infty$ is implicitly understood henceforth.

The electrons are assumed to interact only locally,
and the Hubbard interaction reads
\begin{equation}
\hat{D} = \sum_{\sitei} \left(\hat{n}_{\sitei,\uparrow}-\frac{1}{2}\right)
\left(\hat{n}_{\sitei,\downarrow}-\frac{1}{2}\right) \; ,
\label{defD}
\end{equation}
where $\hat{n}_{\sitei,\sigma}=
\hat{c}^+_{\sitei,\sigma}\hat{c}_{\sitei,\sigma}$ 
is the local density operator at site~$\sitei$ for 
spin~$\sigma$. This 
leads to the Hubbard model~\cite{HubbardI}, 
\begin{equation}
\hat{H}=\hat{T} + U \hat{D} \; .
\label{generalH}
\end{equation}
The Hamiltonian exhibits explicit particle-hole symmetry,
i.e., $\hat{H}$ is invariant under the transformation 
\begin{equation}
{\rm PH}: \qquad \hat{c}_{\sitei,\sigma}^+ \mapsto (-1)^{\sitei}
\hat{c}_{\sitei,\sigma}\quad ; \quad
\hat{c}_{\sitei,\sigma} \mapsto (-1)^{\sitei} \hat{c}_{\sitei,\sigma}^+ \; .
\label{phdef}
\end{equation}
For the Bethe lattice
$(-1)^{\sitei}=+1$ for the $A$~sites which are
surrounded by $B$~sites only (and vice versa), 
for which $(-1)^{\sitei}=-1$. 
A chemical potential $\mu=0$ then guarantees a half-filled band
for all temperatures~\cite{Gebhardbook}.

\subsection{Green functions and self-energy}
\label{subsec:Green}
\label{subsec:selfenetc}

The time-dependent single-particle Green function at zero temperature
is given by~\cite{Fetter}
\begin{equation}
G_{\sigma}(\sitei,\sitej;t) = -\I
\langle \hat{\cal T} [ 
\hat{c}_{\sitei,\sigma}(t)\hat{c}_{\sitej,\sigma}^+] \rangle \; .
\label{GFdef}
\end{equation}
Here $\hat{\cal T}$ is the time-ordering operator, $\langle \ldots \rangle$
implies the expectation value in the ground state with energy $E_0$,
and ($\hbar \equiv 1$)
\begin{equation}
\hat{c}_{\sitei,\sigma}(t) = \exp(\I \hat{H} t) \hat{c}_{\sitei,\sigma}
                          \exp(-\I \hat{H} t) 
\end{equation}
is the annihilation operator in the Heisenberg picture. 
For translationally invariant systems,
the Fourier transform of the Green function is given by
\begin{equation}
G_{\sigma}(\sitek,\omega) = \int_{-\infty}^{\infty} \diff t\, 
e^{\I \omega t} 
\frac{1}{L} \sum_{\sitei,\sitej} e^{-\I \sitek (\sitei-\sitej)} 
G_{\sigma}(\sitei,\sitej;t) \label{DefGkomega} \; .
\end{equation}
The Green function in momentum space can be expressed in terms
of the self-energy $\Sigma_{\sigma}(\sitek,\omega)$~\cite{Fetter} 
($\eta=0^+$),
\begin{equation}
G_{\sigma}(\sitek,\omega) =
\frac{1}{\omega - \epsilon(\sitek)-\Sigma_{\sigma}(\sitek,\omega)
+\I \eta \sgn (\omega)}
\; .
\label{Dyson}
\end{equation}
In the limit of infinite dimensions, the self-energy is independent
of momentum~\cite{MV},
$\Sigma_{\sigma}(\sitek,\omega)=\Sigma_{\sigma}(\omega)$.
As we shall further discuss
in Sect.~\ref{sec:PTU}, the self-energy can be calculated in a power
series in~$U$. 
As shown by Luttinger~\cite{Luttinger},
it has the following properties at small frequency,
\begin{eqnarray}
\Re \Sigma_{\sigma}(\omega)&=& \left(1-\frac{1}{Z}\right) \omega
\quad ; \quad 0\leq \omega < \omega_c 
\label{FLZfactor} \; , \\
\Im \Sigma_{\sigma}(\omega)&=& -\gamma \omega^2 \quad;\quad 0\leq
\omega < \omega_c \; .
\label{FLalpha}
\end{eqnarray}
Here $0<Z\leq 1$, $\gamma\geq 0$, and a low-energy cut-off $\omega_c\ll W$ 
characterize a Fermi liquid.

The local Green function~$G_{\sigma}(\omega)$ 
is the momentum-space average of $G_{\sigma}(\sitek,\omega)$,
\begin{eqnarray}
G_{\sigma}(\omega) &=& \frac{1}{L} \sum_{\sitei} 
\int_{-\infty}^{\infty} \diff t\,  e^{\I \omega t} 
G_{\sigma}(\sitei,\sitei;t) \nonumber \\
& =& \frac{1}{L} \sum_{\sitek} G_{\sigma}(\sitek,\omega)
\label{largedsimplifications}
\\
 &=& \int_{-\infty}^{\infty} \diff\epsilon \rho(\epsilon)
\frac{1}{\omega-\epsilon-\Sigma_{\sigma}(\omega)+\I\eta \sgn (\omega)}
\; .
\nonumber 
\end{eqnarray}
Due to particle-hole symmetry, the local Green function obeys
$G_{\sigma}(t)=-G_{\sigma}(-t)$ so that
$G_{\sigma}(\omega) = - G_{\sigma}(-\omega)$,
and thus $\Sigma_{\sigma}(\omega)=-\Sigma_{\sigma}(-\omega)$.

Because the self-energy only depends on frequency,
the local Green function can be easily recovered from the self-energy.
As seen from~(\ref{largedsimplifications}),
\begin{equation}
G_{\sigma}(\omega) = 
G_{\sigma}^{0}(\omega-\Sigma_{\sigma} (\omega)) \; .
\label{defGlocal}
\end{equation}
This relation is particularly simple
for the semi-circular density of states~(\ref{rhozero}) where
\begin{equation}
G_{\sigma}^{0}(z)=\frac{z}{2}  \left[  
1- \sqrt{1-\frac{4}{z^2}}
\, \right]\; ,
\label{SigmafromG}
\end{equation}
with $z=\omega+\I \eta \sgn (\omega)$.
For the Bethe lattice it then follows that
\begin{equation}
\label{gipt}
\omega-\Sigma_{\sigma}(\omega) =
G_{\sigma}(\omega)+G_{\sigma}^{-1}(\omega)\; .
\end{equation}

With the help of a full set of eigenstates~$|\Psi_n\rangle$,
we can write the local Green function in the Lehmann 
representation~\cite{Fetter}
\begin{eqnarray}
G_{\sigma}(\omega) &=& \int_{-\infty}^{\infty} \diff \omega' 
\frac{D_{\sigma}(\omega')}{\omega-\omega'+\I \eta \sgn(\omega')}
\label{spectralrepresentation}  \\
D_{\sigma}(\omega\geq 0) &=& \sum_{n} \frac{1}{L}\sum_{\sitei}
\left| \langle \Psi_0 | 
\hat{c}_{\sitei,\sigma} | \Psi_n\rangle \right|^2 \delta(\omega+E_0-E_n)
\nonumber \; .
\end{eqnarray}
Consequently, the density of states 
is obtained from the imaginary part of
the local Green function~(\ref{defGlocal}) 
via
\begin{equation}
D_{\sigma}(\omega) = 
- \frac{1}{\pi} \sgn(\omega) \Im G_{\sigma}(\omega)
=D_{\sigma}(-\omega) \; .
\label{Dforlateruse}
\end{equation}
The latter equality holds due to particle-hole symmetry.
For a Fermi liquid, (\ref{FLZfactor}) and (\ref{FLalpha}) 
in~(\ref{defGlocal}) lead to
\begin{equation}
D_{\sigma}(0) = \rho(0)\; ,
\label{pinning}
\end{equation}
i.e., the density of states at $\omega=0$ is pinned to its value
at $U=0$~\cite{MH}.

The moments $M_n$ of the density of states are defined as
\begin{equation}
M_n= 2 \int_{0}^{\infty}
\diff \omega \, \omega^n D_{\sigma}(\omega) \; .
\label{Mndef}
\end{equation}
In particular, from~(\ref{spectralrepresentation}), (\ref{Dforlateruse}),
and the definition of the Green function~(\ref{GFdef}) 
one can show that~\cite{Fetter}
\begin{equation}
M_1=-\frac{1}{L}\left(E_0+U\frac{\partial E_0}{\partial U}\right) 
\label{sumrule2} \; .
\end{equation}
We will later employ this useful sum rule for calculating the
ground-state energy.

The one-particle spectral function $A_{\sigma} (\epsilon;\omega)$
is symmetric in~$\omega$. It is 
defined by ($\omega\geq 0$)
\begin{eqnarray}
A_{\sigma} (\epsilon;\omega) &=& -\frac{1}{\pi} 
\Im\left( \frac{1}{\omega-\epsilon-\Sigma_{\sigma}(\omega)+\I\eta}
\right) 
\label{Defspectral} \\
&=& -\frac{1}{\pi}
\frac{(\Im \Sigma_{\sigma}(\omega)-\eta)}{(\omega-\epsilon-
\Re \Sigma_{\sigma}(\omega))^2+
(\Im \Sigma_{\sigma}(\omega)-\eta)^2} \; .
\nonumber
\end{eqnarray}
Note that 
\begin{equation}
D_{\sigma}(\omega)= \int_{-\infty}^{\infty} \diff\epsilon\, \rho(\epsilon)
A_{\sigma} (\epsilon;\omega) \; .
\end{equation}
As follows from the Lehmann representation~\cite{Fetter},
the spectral function is positive semi-definite,
$A_{\sigma} (\epsilon;\omega) \geq 0$, so that
\begin{equation}
\Im\Sigma_{\sigma}(\omega\geq 0) \leq 0\; .
\label{ImSigmanegative}
\end{equation} 
The spectral function contains a quasi-particle contribution of weight $Z(U)$ 
near the Fermi energy and an incoherent background contribution,
\begin{equation}
A_{\sigma} (\epsilon;\omega\to 0) = Z(U) \delta(\omega-Z(U)\epsilon)
+ A_{\sigma}^{\rm inc} (\epsilon;\omega\to 0) \; .
\end{equation}
The momentum distribution
\begin{equation}
n_{\sigma}(\epsilon) = \int_{-\infty}^{0}
\diff \omega A_{\sigma} (\epsilon;\omega) 
\label{defnofepsilon}
\end{equation}
depends on momentum only implicitly via $\epsilon\equiv\epsilon(\sitek)$.
In the metallic phase, $n_{\sigma}(\epsilon)$
displays
a jump discontinuity at the Fermi energy,
\begin{equation}
n_{\sigma}(\epsilon=0^-)-n_{\sigma}(\epsilon=0^+)= Z(U) \; ,
\label{nepsjump}
\end{equation}
as follows directly from~(\ref{FLZfactor}), (\ref{FLalpha}),
(\ref{defnofepsilon}) and the fact that 
the self-energy does not depend on momentum.
In the vicinity of $\epsilon=0$,
the momentum distribution takes the form ($E \equiv Z(U)\epsilon/\omega_c$)
\begin{equation}
n_{\sigma}(E\ll 1) = \frac{1-Z(U)}{2} 
+ \frac{2\gamma\omega_c Z(U)^2}{\pi}
E\ln E + {\cal O}(E)
\label{epslogeps}
\; .
\end{equation}
This is shown in appendix~\ref{nepsapp}.

\section{Diagrammatic perturbation theory}
\label{sec:PTU}

In this section we derive and calculate the diagrams
to second and fourth order perturbation theory in $U$.

\subsection{Second order}

The particle-hole transformation~(\ref{phdef})
can be restricted to one spin species only. The Hamiltonian
then maps onto itself apart from a change in the sign of~$U$.
Therefore, the Green function obeys
\begin{equation}
G_{\sigma}(\omega;U) = G_{\sigma}(\omega;-U)
\;, 
\end{equation}
and, correspondingly, the self-energy fulfills
\begin{equation}
\Sigma_{\sigma}(\omega;U)=\Sigma_{\sigma}(\omega;-U) \; .
\end{equation}
Consequently, there are no odd orders in the perturbation expansion
of the self-energy in~$U$.

Particle-hole symmetry also guarantees that there are
no (renormalized) Hartree bubbles. A chemical
potential $\mu=0$ results in~\cite{Gebhardbook}
\begin{equation}
\frac{1}{L}\sum_{\sitei}\langle \hat{n}_{\sitei}\rangle
= \frac{1}{L}\sum_{\sitei}\langle \hat{n}_{\sitei}\rangle_0
=\frac{1}{2} \; ,
\end{equation}
i.e., the bare Hartree diagrams are exactly canceled
by definition of the Hubbard interaction in~(\ref{defD}),
and, moreover, the renormalized Hartree diagram vanishes to
all orders in perturbation theory.
Lastly, there are no Fock contributions because the
Hubbard interaction acts between different spin species only.

With these simplifications, only one diagram remains
in second-order perturbation theory, 
\begin{equation}
\Sigma_{-\sigma}^{\rm (2)}(\omega)=
\begin{array}{c}
\resizebox{5cm}{!}{\includegraphics{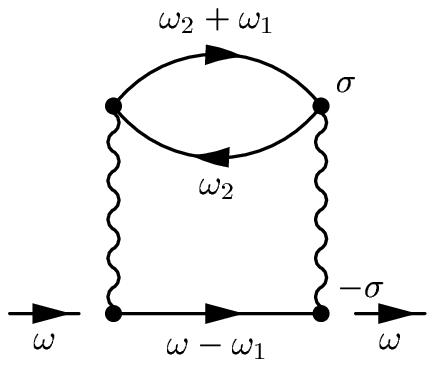}}
\end{array}\; .
\label{2ndorderdiagram}
\end{equation}
Three independent Green function lines connect the two lattice
points. Thus, in the limit of infinite dimensions, 
they can be identified with each other~\cite{MV}, and 
each line thus represents a local bare
Green function~$G_{\sigma}^0(\omega_i)=G_{-\sigma}^0(\omega_i)$
due to spin symmetry.
Note, however, that energy conservation must still be obeyed
at each vertex. Following the Feynman rules, 
the second-order diagram gives the contribution~\cite{MH,DavidSteffen}
\begin{equation}
\Sigma_{-\sigma}^{\rm (2)}(\omega)
=(-1)(\I)^2 U^2\int_{-\infty}^{\infty}\frac{\diff\omega_1}{2\pi\I}
G_{-\sigma}^0(\omega-\omega_1)\Pi_{\sigma}^0(\omega_1)
\label{Sig2ndFeyn}
\end{equation} 
with the bare polarization bubble
\begin{equation}
\Pi_{\sigma}^0(\omega)
= - \int_{-\infty}^{\infty}\frac{\diff\omega_2}{2\pi\I}\,
G_{\sigma}^0(\omega_2)
G_{\sigma}^0(\omega_2+\omega)
= \Pi_{\sigma}^0(-\omega) \; .
\end{equation} 
With the help of the spectral representation~(\ref{largedsimplifications})
of the local bare Green function
\begin{equation}
\label{G0}
G_{\sigma}^0(\omega)=\int_0^{W/2}\diff\epsilon\, \rho(\epsilon)
\left(\frac{1}{\omega-\epsilon+\I \eta}
+\frac{1}{\omega+\epsilon-\I\eta}\right) \; ,
\end{equation}
a contour integration results in
\begin{eqnarray}
\Pi_{\sigma}^0(\omega)&=&-\int_0^{W/2}\diff\epsilon_1
\int_0^{W/2}\diff\epsilon_2\, \rho(\epsilon_1) \rho(\epsilon_2)
\label{fullPi}
\\
&& \left(\frac{1}{\omega-\epsilon_{1}-\epsilon_{2}+\I \eta}
-\frac{1}{\omega+\epsilon_{1}+\epsilon_{2}-\I \eta}
\right) \; .
\nonumber 
\end{eqnarray}
Taking the imaginary part gives
\begin{equation}
\label{imagpi}
\frac{1}{\pi}\Im\Pi_{\sigma}^0(\omega\geq 0) 
=\int_0^{\omega}\diff\epsilon\,\rho(\epsilon)
\rho(\omega-\epsilon)\; .
\end{equation}
This representation explicitly shows that
the imaginary part of the bare polarization bubble
vanishes for $|\omega|\geq W$. This is a consequence of the
fact that the bare polarization bubble is made up of
two bare Green function lines.

Using the spectral representation of the bare polarization
bubble~(\ref{fullPi}) and of the local Green function~(\ref{G0})
in~(\ref{Sig2ndFeyn}), the contour integration over $\omega_1$
can easily be performed. Taking the imaginary part leads to 
\begin{equation}
\label{s2}
\Im\Sigma_{-\sigma}^{\rm (2)}(\omega\geq 0)
=-U^2\int_0^{\omega}\diff\epsilon\,\rho(\epsilon)
\Im\Pi_{\sigma}^0(\omega-\epsilon) \; .
\label{ImS2result}
\end{equation}
The imaginary part of the 
second-order self-energy vanishes for $|\omega|\geq 3W/2$.
The Hilbert transformation provides the real part as
\begin{equation}
\Re\Sigma_{-\sigma}^{\rm (2)}(\omega)=
\frac{1}{\pi}{\cal P}\int_0^{3W/2}\diff\zeta\,
\Im\Sigma_{-\sigma}^{\rm (2)}(\zeta) \left( \frac{1}{\zeta-\omega}
- \frac{1}{\zeta+\omega}\right)\; .
\end{equation}
For practical calculations it is advisable to split
the integration routines into intervals $[(r-1) W/2,r W/2]$
($r=1,2,3$) in order to speed up the integrations and to
minimize numerical errors.

\begin{figure}[htb]
 \resizebox{8cm}{!}{\includegraphics{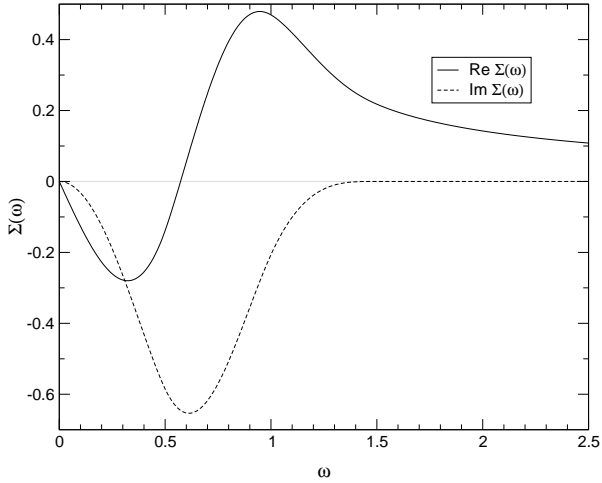}}
\caption{Real and imaginary part of the second-order self-energy;
units $W=4t\equiv 1$.\label{zweiteOrdnung}}
\end{figure}

For the density of states of the Bethe lattice~(\ref{rhozero}),
the result for the real and imaginary parts of the second-order
self-energy are shown in Fig.~\ref{zweiteOrdnung}, apart from
the prefactor $U^2$. 
As seen from the figure, the self-energy reproduces the
Fer\-mi-liquid relations~(\ref{FLZfactor}) 
and~(\ref{FLalpha}) for small frequencies. Explicitly, we find
from our numerical integrations
\begin{equation}
Z(U)^{-1} = 1+ 1.307[1] \left(\frac{U}{W}\right)^2 
+ {\cal O}(U^4)
\end{equation}
and 
\begin{equation}
\gamma(U) = 3.242[1] \left(\frac{U}{W}\right)^2 + {\cal O}(U^4) \; ,
\end{equation}
where the number in brackets denotes the uncertainty in the last digit.

\subsection{Fourth order}

The twelve topologically different diagrams to fourth order 
can be grouped into four sets; see below. They 
were used earlier
by Yamada and Yosida~\cite{YY} in their study of
the symmetric Anderson impurity model, and 
by Freericks and Jarrell~\cite{FreeJarrell}
in their finite-temperature perturbation study of the Hubbard model.
With the help of particle-hole symmetry it is not difficult
to show that the diagrams of each set give the same 
contribution~\cite{FreeJarrell}.
We also find
\begin{equation}
\Sigma_{\sigma}^{\rm (4)}(\omega) = 3 \biggl(  
\Sigma_{\sigma}^{\rm (4a)}(\omega)+ 
\Sigma_{\sigma}^{\rm (4b)}(\omega)+
\Sigma_{\sigma}^{\rm (4c)}(\omega)+
\Sigma_{\sigma}^{\rm (4d)}(\omega) \biggr) \; .
\end{equation}
We discuss the diagrams of the four sets and their
contributions to the self-energy in the following.

\subsubsection{Ring diagram}

\begin{figure}[htb]
$3 \Sigma_{\sigma}^{\rm (4a)}(\omega)\quad=\qquad
\begin{array}{c}
 \resizebox{3cm}{!}{\includegraphics{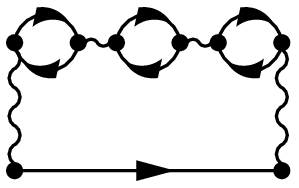}}\\[12pt]
+ \\[30pt]
\resizebox{3cm}{!}{\includegraphics{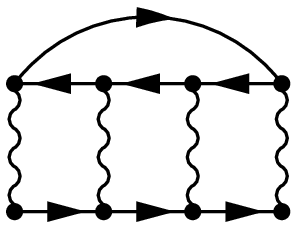}}\\[12pt]
+ \\[30pt]
\resizebox{3cm}{!}{\includegraphics{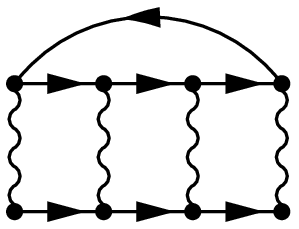}}
\end{array}$
\caption{Set A of three equivalent diagrams to the proper self-energy
in fourth-order perturbation theory.\label{Fig:4thorderdiagramA}}
\end{figure}

The ring diagram in Fig.~\ref{Fig:4thorderdiagramA} 
and the particle-hole/par\-ti\-cle-particle ladders give identical contributions at
half band-filling. This is most easily seen when the Feynman rules
are applied in the time domain, and particle-hole symmetry,
$G_{\sigma}(t)=-G_{\sigma}(-t)$, is used appropriately.
For the ring diagram the Feynman rules result in
\begin{equation}
\Sigma_{\sigma}^{\rm (4a)}(\omega)=
\begin{array}{c}
\resizebox{6cm}{!}{\includegraphics{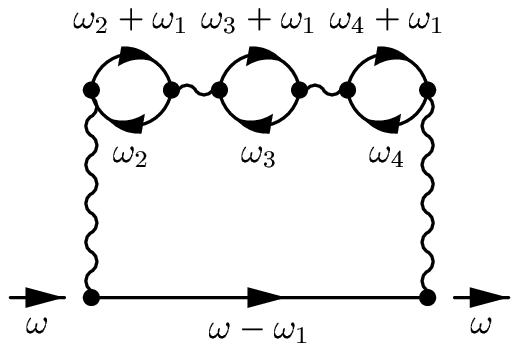}}
\end{array}\; .
\end{equation}
In order to evaluate this diagram, we define
\begin{equation}
P(\omega)=\left[\Pi_{\sigma}^0(\omega)\right]^{3} \; .
\end{equation}
Analogously to the second-order calculation, we then find
\begin{equation}
\Im\Sigma_{\sigma}^{\rm (4a)}(\omega\geq 0)
=-U^4\int_0^{\omega}
\diff\epsilon\,\rho(\epsilon)\Im P(\omega-\epsilon) 
\end{equation}
with
\begin{equation}
\Im P(\omega)=\Im\Pi_{\sigma}^0(\omega)
\left[3(\Re\Pi_{\sigma}^0(\omega))^2-(\Im\Pi_{\sigma}^0(\omega))^2\right]
\; .
\end{equation}
The real part $\Re\Pi_{\sigma}^0(\omega)$ is obtained
via Hilbert transformation of $\Im\Pi_{\sigma}^0(\omega)$;
see appendix~\ref{appB}.

\subsubsection{Second-order diagram with second-order vertex correction}

\begin{figure}[htb]
$3 \Sigma_{\sigma}^{\rm (4b)}(\omega)\quad=\qquad
\begin{array}{c}
 \resizebox{3cm}{!}{\includegraphics{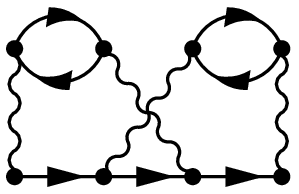}}\\[12pt]
+ \\[30pt]
\resizebox{3cm}{!}{\includegraphics{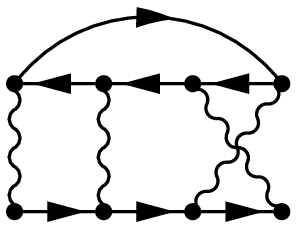}}\\[12pt]
+ \\[30pt]
\resizebox{3cm}{!}{\includegraphics{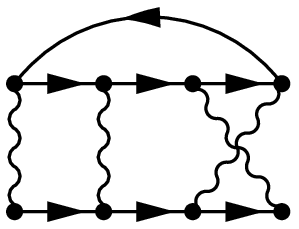}}
\end{array}$
\caption{Set B of three equivalent diagrams to the proper self-energy
in fourth-order perturbation theory.\label{Fig:4thorderdiagramB}}
\end{figure}

The second-order diagram with second-order vertex correction
in Fig.~\ref{Fig:4thorderdiagramB} and
the particle-hole/particle-particle ladders
with crossed interaction lines give identical contributions at
half band-filling. 
For the second-order diagram with second-order vertex correction,
the Feynman rules result in
\begin{equation}
\Sigma^{\rm (4b)}(\omega)=
\begin{array}{c}
\resizebox{6.65cm}{!}{\includegraphics{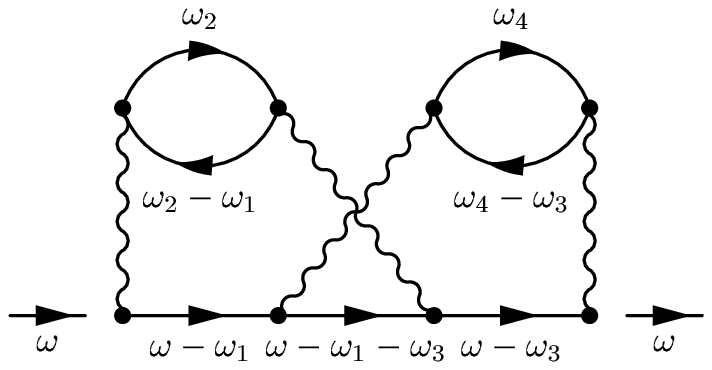}}
\end{array}\; .
\end{equation}
Explicitly, 
\begin{eqnarray}
\Sigma_{\sigma}^{\rm (4b)}(\omega)&=&U^4
\int_{-\infty}^{\infty}\frac{\diff\omega_1}{2\pi\I}
\int_{-\infty}^{\infty}\frac{\diff\omega_3}{2\pi\I}
G_{\sigma}^0(\omega-\omega_1)\Pi_{\sigma}^0(\omega_1)
 \nonumber\\
&& \hphantom{-U^4}
\times 
G_{\sigma}^0(\omega-\omega_1-\omega_3)G_{\sigma}^0(\omega-\omega_3)
\Pi_{\sigma}^0(\omega_3) \; .
 \nonumber\\
&& \label{Sigma4bFeyn}
\end{eqnarray}

The contour integration over $\omega_3$
can be performed using the spectral representation of the bare polarization
bubble~(\ref{fullPi}) and of the local Green function~(\ref{G0})
in~(\ref{Sigma4bFeyn}). 
The remaining contour integral
over $\omega_1$ then results in 24~terms, four of which are zero.
When we restrict ourselves to $\omega>0$ and focus on the
imaginary part, another four terms vanish. 
A change in integration variables shows that six terms
come in pairs. Thus, only ten different terms need to be 
evaluated~\cite{Sandradiss}. The calculation of the imaginary
part reduces the seven-fold integrations over the
bare density of states to six-fold integrations. Fortunately,
these integrations can be grouped so that the final result
can be expressed in terms of help functions which can be tabulated
and used in the remaining at most two-fold integrations
over finite energy intervals.
These help functions are listed in appendix~\ref{appB}.

The ten remaining terms lead to 20~integrals which can be
expressed using the bare density of states and the help
functions. Appropriate changes in the energy integration variables
allow us to regroup them into six terms,
\begin{equation}
\Im\Sigma_{\sigma}^{\rm (4b)}(\omega\geq 0) = \pi U^4
\sum_{i=1}^6 I_i
\end{equation}
with
\begin{eqnarray}
I_1 &=&-\int_0^W \diff a\, h(a)\int_0^W \diff b\, h(b)\nonumber \\
&& \hphantom{-} 
\widetilde{\rho}(\omega-a-b) l(b-\omega) l(a-\omega)\; ,\\[3pt]
I_2 &=&-2\int_0^W\diff b\, h(b)\int_0^{W/2}\diff\epsilon_1
\rho(\epsilon_1)h(\omega-\epsilon_1) \nonumber\\
&& \hphantom{-2} 
\left[f(b-\omega)f(b-\epsilon_1)+f(b+\omega)f(b+\epsilon_1)\right]
\; ,\\[3pt]
I_3&=&2\int_0^W\diff a\,
h(a)\widetilde{\rho}(\omega-a)\int_0^{W/2}\diff\epsilon_1
\rho(\epsilon_1)f(\epsilon_1+a)\nonumber \\ 
&& \hphantom{2} 
\left[H(\epsilon_1+a-\omega)+H(\omega+\epsilon_1)\right]
\; ,\\[3pt]
I_4 &=&\pi^2\int_0^W\diff a\, h(a)\int_0^W\diff b\, 
h(b)\widetilde{\rho}(\omega-a)
\widetilde{\rho}(\omega-b)\nonumber \\ 
&& \hphantom{\pi^2} 
\left[\widetilde{\rho}(\omega-a-b)-\widetilde{\rho}(a+b-\omega)\right] 
\; ,\\[3pt]
I_5&=&2\int_0^{W/2}\diff\epsilon_1\rho(\epsilon_1)
\int_0^{W/2}\diff\epsilon_2\rho(\epsilon_2)
h(\omega-\epsilon_1)
\nonumber\\ 
&& \hphantom{2} 
f(\epsilon_1+\epsilon_2-\omega)
\left[H(\epsilon_1+\epsilon_2)+H(\epsilon_2-\omega)\right]
\;, \\[3pt]
I_6 &=&\int_0^{W/2}\diff\epsilon_1\rho(\epsilon_1)
\int_0^{W/2}\diff\epsilon_2\rho(\epsilon_2)
\widetilde{\rho}(\omega-\epsilon_1-\epsilon_2) \nonumber \\ 
&&\Bigl[2H(\epsilon_2-\omega)H(\epsilon_1+\epsilon_2)
+H(\omega-\epsilon_2)H(\epsilon_1+\epsilon_2) \nonumber \\
&& \hphantom{\Bigl[}
+H(\epsilon_1-\omega)H(\epsilon_2-\omega)\Bigr] 
\; ,
\end{eqnarray} 
where $\widetilde{\rho}(x)=\rho(x)\Theta(x)$.
For practical calculations, it is advisable to split
the integration routines into intervals $[(r-1) W/2,r W/2]$
($r=1,\ldots,5$) in order to speed up the integrations and to
minimize numerical errors.
We have checked our results against a numerical integration
in the time domain which is easier to implement but much less accurate
for the same computational effort.

\begin{figure}[htb]
$3 \Sigma_{\sigma}^{\rm (4c)}(\omega)\quad=\qquad
\begin{array}{c}
 \resizebox{3cm}{!}{\includegraphics{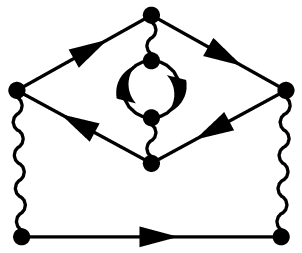}}\\[12pt]
+ \\[30pt]
\resizebox{3cm}{!}{\includegraphics{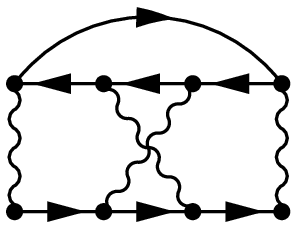}}\\[12pt]
+ \\[30pt]
\resizebox{3cm}{!}{\includegraphics{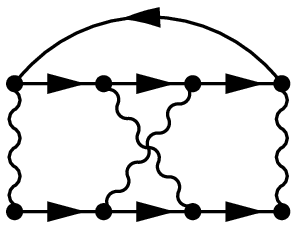}}
\end{array}$
\caption{Set C of three equivalent diagrams to the proper self-energy
in fourth-order perturbation theory.\label{Fig:4thorderdiagramC}}
\end{figure}

\subsubsection{Second-order diagram with vertex correction
in the polarization bubble}

The second-order diagram with vertex correction in the polarization bubble
in Fig.~\ref{Fig:4thorderdiagramC} and  
the par\-ticle-hole/par\-ti\-cle-particle ladders
with crossed interaction lines give identical contributions at
half band-filling. 
For the second-order diagram with second-order vertex corrections
in the polarization bubble, the Feynman rules result in
\begin{equation}
\Sigma_{\sigma}^{\rm (4c)}(\omega)=
\begin{array}{c}
\resizebox{6.65cm}{!}{\includegraphics{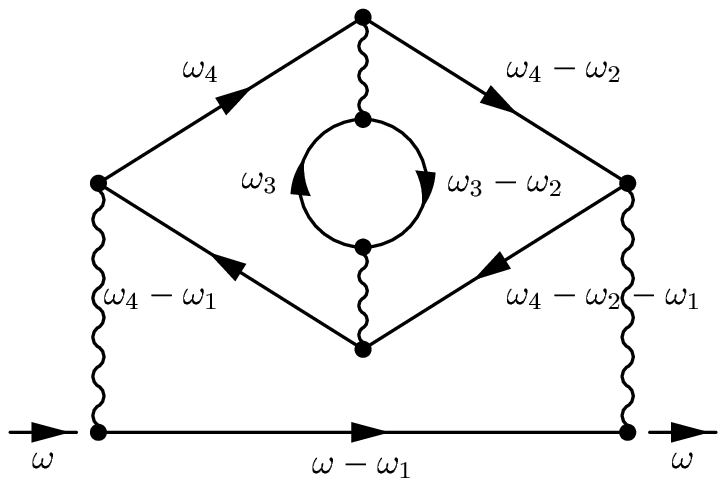}}
\end{array} \; .
\end{equation}
Obviously, we may split off the renormalized polarization bubble.
Analogously to the second-order calculation, we immediately arrive at
\begin{equation}
\Im \Sigma_{\sigma}^{\rm (4c)}(\omega\geq 0)=-U^4\int_0^{\omega}\diff\epsilon
\rho(\epsilon) \Im\Pi_{\rm V}(\omega-\epsilon)
\end{equation}
with 
\begin{eqnarray}
\Pi_{\rm V}(x)&=&
- \int_{-\infty}^{\infty}\frac{\diff\omega_2}{2\pi\I}
\int_{-\infty}^{\infty}\frac{\diff\omega_4}{2\pi\I} 
 G_{\sigma}^0(\omega_4) G_{\sigma}^0(\omega_4-x) 
\nonumber 
\\
&&
G_{\sigma}^0(\omega_4-\omega_2)
G_{\sigma}^0(\omega_4-x-\omega_2)
\Pi_{\sigma}^0(\omega_2)
\;,\label{renbubble}
\end{eqnarray}
which is symmetric in $x$.

Using the spectral representation of the bare polarization
bubble~(\ref{fullPi}) and of the local Green function~(\ref{G0})
in~(\ref{renbubble}), the contour integration over $\omega_2$
results in six non-vanishing terms. The remaining contour integral
over $\omega_4$ then gives 24~terms of which seven are zero,
and four come in pairs. Of the remaining 13~terms, five do not
provide a finite imaginary part for $x\geq 0$.
A re-grouping of integration variables shows that only five
different terms remain for $x\geq 0$,
\begin{eqnarray}
\Im\Pi_{\rm V}(x)\!&=&\!
2\Im \int_0^{\infty}\diff\epsilon_1
\rho(\epsilon_1)\dots\int_0^{\infty}\diff\epsilon_6\rho(\epsilon_6)
\nonumber 
\\
%
%
&&\biggl\{
\frac{1}{x+\epsilon_1+\epsilon_3+\epsilon_4+\epsilon_6}
\frac{1}{\epsilon_1+\epsilon_3+\epsilon_4+\epsilon_5} \nonumber\\
&& \hphantom{\int_0^{\infty}\diff\epsilon_1}
\times
\frac{1}{x+\epsilon_1-\epsilon_2-\I\eta_1\sgn(\epsilon_1-\epsilon_2)}
\nonumber\\[3pt]
%
%
&&+\frac{1}{x-\epsilon_2-\epsilon_3-\epsilon_4-\epsilon_5+\I\eta_3}
\frac{1}{\epsilon_2+\epsilon_3+\epsilon_4+\epsilon_6} \nonumber\\
&& \hphantom{\int_0^{\infty}\diff\epsilon_1}
\times
\frac{1}{x+\epsilon_1-\epsilon_2-\I\eta_1\sgn(\epsilon_1-\epsilon_2)}
\nonumber\\[3pt]
%
%
&&+2\frac{1}{x-\epsilon_1-\epsilon_2+\I\eta_1}
\frac{1}{x-\epsilon_2-\epsilon_3-\epsilon_4-\epsilon_5+\I\eta_5}\nonumber
\\
&& \hphantom{\int_0^{\infty}\diff\epsilon_1}
\times
\frac{1}{\epsilon_2+\epsilon_3+\epsilon_4+\epsilon_6}
\nonumber\\[3pt]
%
%
&&+2\frac{1}{x-\epsilon_1-\epsilon_2+\I\eta_1}
\frac{1}{x+\epsilon_5+\epsilon_6-\I\eta_2} \nonumber \\
&& \hphantom{\int_0^{\infty}\diff\epsilon_1}
\times
\frac{1}{\epsilon_2+\epsilon_3+\epsilon_4+\epsilon_6}
\nonumber\\[3pt]
%
%
&&+\frac{1}{x-\epsilon_1-\epsilon_2+\I\eta_1}
\frac{1}{x-\epsilon_5-\epsilon_6+\I\eta_2} \nonumber \\
&& \hphantom{\int_0^{\infty}\diff\epsilon_1}
\times
\frac{1}{x-\epsilon_2-\epsilon_3-\epsilon_4-\epsilon_5+\I\eta_3}
\biggr\}
\;.\label{Pi_v.1}
\end{eqnarray}
Taking the imaginary part simplifies this expression.
First, the terms with the common factor $\sgn(\epsilon_{1}-\epsilon_{2})
\delta(x+\epsilon_{1}-\epsilon_{2})$ cancel each other. 
Second, we are left with five-fold integrations over 
finite intervals. Third, these expressions can be simplified
further using the help functions of appendix~\ref{appB}. 
We find
\begin{equation}
\Im\Pi_{\rm V}(x\geq 0) = -2\pi 
\sum_{i=1}^7 J_i
\end{equation}
with
\begin{eqnarray}
J_1&=&\int_0^{W/2}\diff\epsilon_2\int_0^W\diff y
\nonumber \\ &&
h(y)\rho(\epsilon_2)
\widetilde{\rho}(x-\epsilon_2-y)f(x-\epsilon_2)
f(\epsilon_2+y)
\, , 
\\ 
%
J_2&=& -2\int_0^{W/2}\diff\epsilon_2\int_0^W\diff y
\nonumber \\ &&
h(y)\rho(\epsilon_2)\widetilde{\rho}(x-\epsilon_2)
f(\epsilon_2+y-x)f(\epsilon_2+y)
 \, , \\ 
%
J_3 &=&-2\int_0^{W/2}\diff\epsilon_2\int_0^W\diff y
\nonumber \\ &&
h(y)\rho(\epsilon_2)\widetilde{\rho}(x-\epsilon_2-y)
f(\epsilon_2-x)f(\epsilon_2+y)
\, , \\ 
%
J_4 &=& 2\int_0^{W/2}\diff\epsilon_2\int_0^{W/2}\diff\epsilon_6
\nonumber \\ &&
\rho(\epsilon_2)\rho(\epsilon_6)\widetilde{\rho}(x-\epsilon_2)
f(x+\epsilon_6)H(\epsilon_2+\epsilon_6)
\, , \\ 
%
J_5&=& 2\int_0^{W/2}\diff\epsilon_2\int_0^{W/2}\diff\epsilon_5
\nonumber \\ &&
\rho(\epsilon_2)\rho(\epsilon_5)\widetilde{\rho}(x-\epsilon_2)
f(\epsilon_5-x)H(\epsilon_2+\epsilon_5-x)
 \nonumber \, , \\ && \\
J_6&=&\int_0^{W/2}\diff\epsilon_2\int_0^{W/2}\diff\epsilon_5
\nonumber \\ &&
\rho(\epsilon_2)\rho(\epsilon_5)f(\epsilon_2-x)
f(\epsilon_5-x)h(x-\epsilon_2-\epsilon_5)
\, , \\ 
%
J_7 &=& -\pi^2\int_0^{W/2}\diff\epsilon_2\int_0^{W/2}\diff\epsilon_5
\nonumber \\ &&
\rho(\epsilon_2)\rho(\epsilon_5)\widetilde{\rho}(x-\epsilon_{2})
\widetilde{\rho}(x-\epsilon_5)h(x-\epsilon_2-\epsilon_5)
\label{Pi_v.2} \, , 
%
\end{eqnarray}
where $\widetilde{\rho}(x)=\rho(x)\Theta(x)$.
For practical calculations it is advisable to split
the integration routines into intervals $[(r-1) W/2,r W/2]$
($r=1,\ldots,4$) in order to speed up the integrations and to
minimize numerical errors.
We have checked our results against a numerical integration
in the time domain.

\subsubsection{Second-order diagram with second-order
self-energy insertion}

\begin{figure}[hbt]
$3 \Sigma_{\sigma}^{\rm (4d)}(\omega)\quad=\qquad
\begin{array}{c}
 \resizebox{3.5cm}{!}{\includegraphics{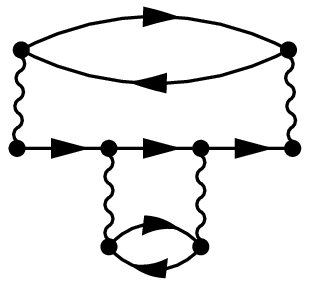}}\\[12pt]
+ \\[18pt]
\resizebox{3.5cm}{!}{\includegraphics{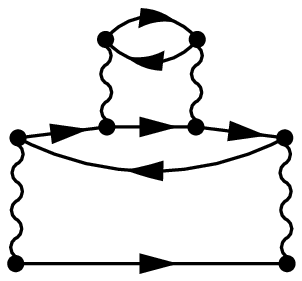}}\\[12pt]
+ \\[18pt]
\resizebox{3.5cm}{!}{\includegraphics{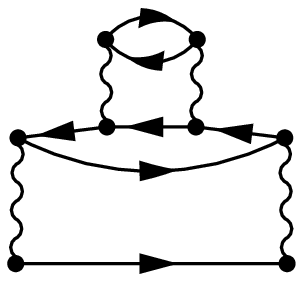}}
\end{array}$
\caption{Set D of three equivalent diagrams to the proper self-energy
in fourth-order perturbation theory.\label{Fig:4thorderdiagramD}}
\end{figure}

The second-order diagrams with self-energy insertion
in Fig.~\ref{Fig:4thorderdiagramD} 
give identical contributions in the paramagnet
at half band-filling. 
These diagrams are not skeleton diagrams so that momentum conservation
cannot be ignored at the inner vertices.

For the second-order diagram with self-energy insertion, 
the Feynman rules result in
\begin{equation}
\Sigma_{\sigma}^{\rm (4d)}(\omega)=
\begin{array}{c}
\resizebox{6cm}{!}{\includegraphics{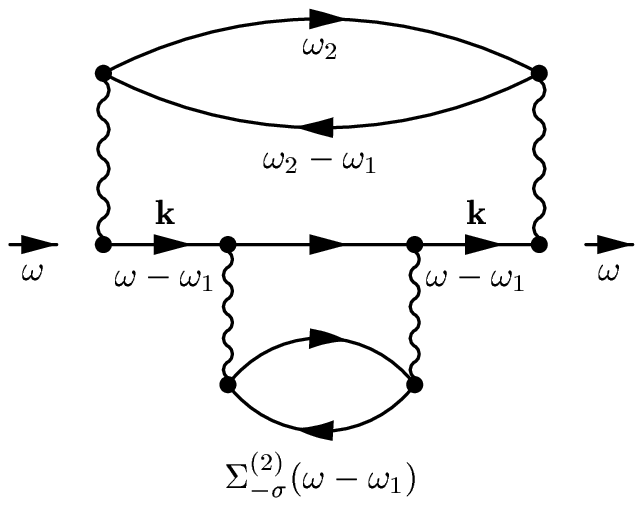}}
\end{array} \; .
\label{eqS4d}
\end{equation}
Obviously, we may use the results for the bare polarization bubble
and the second-order self-energy to simplify the expression to the form
\begin{eqnarray}
\Sigma_{\sigma}^{\rm (4d)}(\omega)&=&
U^2\int_{-\infty}^{\infty}\frac{\diff\omega_{1}}{2\pi\I}
\Pi_{\sigma}^0(\omega_1) \Sigma_{\sigma}^{(2)}(\omega-\omega_1) 
\nonumber \\
&& \hphantom{U^2}
\frac{1}{L}\sum_{\sitek} [G_{\sigma}^0(\epsilon(\sitek);\omega-\omega_1)]^2
\; .
\label{S4dequation}
\end{eqnarray}
We must evaluate
\begin{equation}
\frac{1}{L}\sum_{\sitek}
     [G_{\sigma}^0(\epsilon(\sitek);x)]^2
= A_+(x)+A_-(x)
\label{G0kepsilon)} 
\; , 
\end{equation}
with
\begin{equation}
A_{\pm}(x) = \int_0^{W/2}\diff y \rho(y)  
\frac{1}{x\mp y\pm \I\eta_3} \frac{1}{x\mp y\pm \I\eta_4} \; .
\label{Ayspectral}
\end{equation}
We may set $\eta_3=\eta_4$ later so that 
$A_{\pm}(x)$ may be expressed in terms
of $\rho(0)$ and the derivative of the bare density of states,
$d(x)=(\diff \rho(x))/(\diff x)$.

The spectral representations of $A_{\pm}(x)$ in~(\ref{Ayspectral}),
$\Pi_{\sigma}^0(\omega)$ in~(\ref{spectralpi}), and
$\Sigma_{\sigma}^{(2)}(\omega)$ in~(\ref{spectrals})
allow us to perform the contour integral over $\omega_1$
in~(\ref{S4dequation}) with the result
\begin{eqnarray}
\Im \Sigma_{\sigma}^{\rm (4d)}(\omega)&=&\pi U^4
\int\limits_0^{W/2}\diff x\, 
d(x) h(\omega-x)[S(x)-S(-x)]\nonumber \\
&& +\pi U^4\int_0^{3W/2}\diff b\,
s(b)h(\omega-b)l^{\prime}(b) \; ,
\end{eqnarray}
where $l^{\prime}(b)$ denotes the derivative of the help function
$l(b)$, see~(\ref{l}).
Again, it is advisable to split
the integration routines into intervals $[(r-1) W/2,r W/2]$
($r=1,\ldots,5$) in order to speed up the integrations and to
minimize numerical errors.

\begin{figure}[htb]
 \resizebox{8cm}{!}{\includegraphics{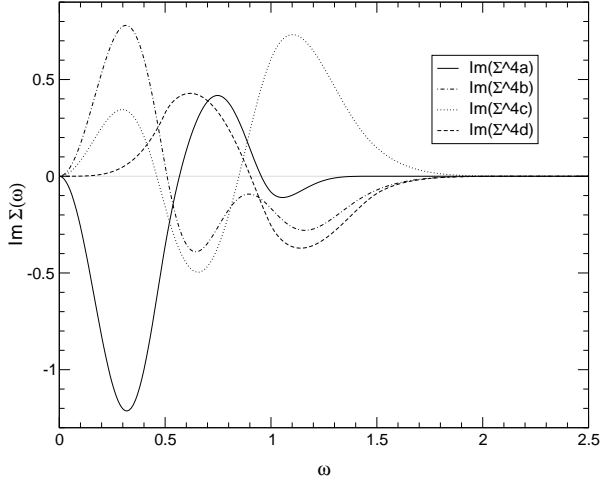}}
\caption{Imaginary part of the fourth-order self-energy diagrams;
units $W=4t\equiv 1$.\label{vierteOrdnungDiags}}
\end{figure}

The result for the real and imaginary parts of the fourth-order
diagrams are shown in Fig.~\ref{vierteOrdnungDiags}, apart from
the prefactor $U^4$. 
As seen from the figure, the diagrams are equally important.
At half band-filling there is generally no reason to include
only special diagram classes, as done, e.g., in the RPA or in the ladder
approximation. In particular, for small $\omega$ the 
contribution of the fourth-order ring diagram is to a large
extent canceled by the two diagrams with vertex corrections.

\begin{figure}[htb]
 \resizebox{8cm}{!}{\includegraphics{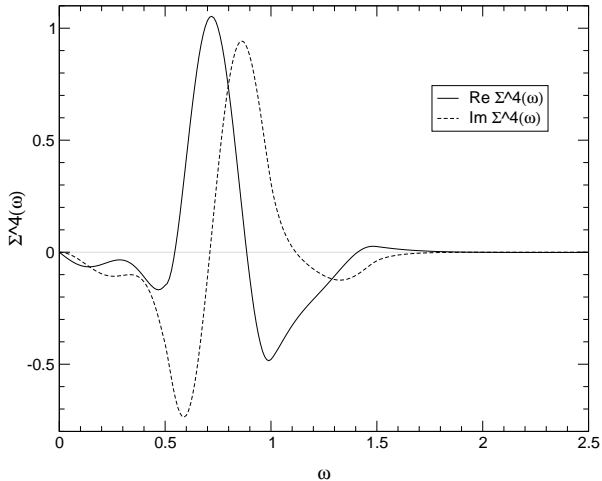}}
\caption{Real and imaginary part of the fourth-order self-energy;
units $W=4t\equiv 1$.\label{vierteOrdnung}}
\end{figure}

Fig.~\ref{vierteOrdnung} shows the real and imaginary parts of the fourth-order
self-energy, apart from the prefactor $U^4$. 
The self-energy reproduces the
Fermi-liquid relations~(\ref{FLZfactor}) 
and~(\ref{FLalpha}) for small frequencies. Explicitly, we find
from our numerical integrations 
\begin{equation}
Z(U)^{-1} = 1+ 1.307[1] \left(\frac{U}{W}\right)^2 
+ 0.739[1] \left(\frac{U}{W}\right)^4 + {\cal O}(U^6)
\label{4thZ}
\end{equation}
and 
\begin{equation}
\gamma(U) = 3.242[1] \left(\frac{U}{W}\right)^2 
+ 4.22[1] \left(\frac{U}{W}\right)^4 + {\cal O}(U^6) \; ,
\end{equation}
where the number in brackets denotes the uncertainty in the last digit.
The prefactor of the fourth-order term in
$[Z(U)]^{-1}$ is smaller than reported in~\cite{RDA}.
There, the momentum distribution was calculated using 
Rayleigh-Schr\"odinger stationary perturbation theory.
We suspect that stationary perturbation theory is flawed when
diagrams with self-energy insertions as in~(\ref{eqS4d}) occur.

Fig.~\ref{vierteOrdnung} shows that the fourth-order contribution
to the self-energy becomes {\em positive} for $0.7 W \leq \omega \leq 1.1W$.
For small interaction strengths this positive contribution
is compensated by the overall negative second-order self-energy.
The sum of both terms is no longer negative for all
frequencies above $U=0.64 W$, in contrast
to the exact result~(\ref{ImSigmanegative}).
This limits the applicability of fourth-order perturbation
theory to moderate interactions strengths.

\subsection{Ground-state energy and average double occupancy}

The ground-state energy can also be 
expanded in terms of powers in $(U/W)^2$,
\begin{equation}
\frac{E_0}{L} = -\frac{2W}{3\pi} + \alpha W
\left(\frac{U}{W}\right)^2 + \beta W \left(\frac{U}{W}\right)^4 +
{\cal O}(U^6) \; .
\end{equation}
We use this expansion in~(\ref{sumrule2}) and obtain
\begin{equation}
M_1 = \frac{2W}{3\pi} - 3 \alpha W \left(\frac{U}{W}\right)^2 
- 5 \beta W \left(\frac{U}{W}\right)^4 + {\cal O}(U^6) \; .
\end{equation}
On the other hand, we may use~(\ref{Mndef}) into which we
insert the density of states from~(\ref{Dforlateruse})
which we can express in terms of the self-energy with the help
of~(\ref{defGlocal}) and~(\ref{SigmafromG}). For a better numerical
accuracy, we calculate
\begin{eqnarray}
a &=& \int_0^{3W/2} \diff\omega\, \omega 
\Im\Bigl[ -\Sigma_{\sigma}^{(2)}(\omega) 
-\sqrt{\bigl[\omega-\Sigma_{\sigma}^{(2)}(\omega) \bigr]^2-4} 
\nonumber \\
&& 
\hphantom{ \int_0}
 +\sqrt{\omega^2-4}\, \Bigr] \; ,\\
b &=& \int_0^{5W/2} \diff\omega\, \omega 
\Im\Bigl[ -\Sigma_{\sigma}^{(4)}(\omega) 
+\sqrt{\bigl[\omega-\Sigma_{\sigma}^{(2)}(\omega)\bigr]^2-4}
\nonumber  \\
&& 
\hphantom{\int_0}
+\Sigma_{\sigma}^{(2)}(\omega) 
-\sqrt{\bigl[\omega-\Sigma_{\sigma}^{(2)}(\omega)
-\Sigma_{\sigma}^{(4)}(\omega)\bigr]^2-4}
\Bigr] 
\; . \nonumber \\
&& 
\end{eqnarray}
We fit the results to the form $a=a_0 (U/W)^2 + a_1 (U/W)^4$,
and $b=b_0 (U/W)^4 + b_1 (U/W)^6$, and obtain
for $\alpha=a_0/(3\pi)$ and $\beta=b_0/(5\pi)$
\begin{eqnarray}
\alpha&=& - 0.08346[1] \; , \\
\beta&=& - 0.0062[2] \; .
\end{eqnarray}
The prefactor of the fourth-order term, $\beta$, 
is an order of magnitude smaller 
than the second-order prefactor, $\alpha$. This is very similar to the result
for the single-impurity Anderson model addressed by 
Yamada and Yosida~\cite{YY}. Therefore, we expect that the ground-state energy
is reasonably well-described by fourth-order perturbation theory 
up to $U\leq W$.

Starting from 
\begin{equation}
\overline{d}(U) = \frac{1}{4} + \frac{1}{L} \langle \hat{D} \rangle  
= \frac{1}{4} + \frac{1}{L} \frac{\partial E_0}{\partial U}
\end{equation}
we obtain the fourth-order result
\begin{eqnarray}
\overline{d}(U) &=& \frac{1}{4} + 2\alpha \left(\frac{U}{W}\right) 
+ 4 \beta \left(\frac{U}{W}\right)^3 + {\cal O}(U^5) \nonumber \\
&=& 0.25 - 0.16692[2] \left(\frac{U}{W}\right)
- 0.0248[8] \left(\frac{U}{W}\right)^3 \label{dbar4}\\ 
&& + {\cal O}(U^5)\nonumber 
\end{eqnarray}
for the average double occupancy at half band-filling.
The average double occupancy is positive semi-definite. This criterion
excludes the applicability of~(\ref{dbar4}) for interaction
strengths $U\geq 1.12[1] W$.

An even more stringent condition can be drawn from a comparison with
the ground-state energy of the Mott-Hubbard insulator at half 
band-filling~\cite{EastwoodGebhard}. Up to third-order in $1/U$,
\begin{equation}
\frac{E_0}{L} = -\frac{U}{4} -\frac{W^2}{32 U}- \frac{W^4}{512 U^3} 
+ {\cal O}(U^{-5}) \; ,
\label{E01overU}
\end{equation}
and
\begin{equation}
\overline{d}(U) = \frac{1}{32} \left(\frac{W}{U}\right)^2 
+ \frac{3}{512} \left(\frac{W}{U}\right)^4 
+ {\cal O}(U^{-6}) \; .
\end{equation}
If we use the $1/U$-expansion down to $U\geq W$, we find
that the average double occupancy from the expansion in~$U$ and
in $1/U$ become equal at the crossing point $U_{\rm cross}=1.056 W$.
Note that in~\cite{EastwoodGebhard} we estimated $U_{{\rm c},1}=1.105[10]W$ 
for the opening of the gap. The fact that $U_{\rm cross}\approx
U_{{\rm c},1}$ might indicate that the Mott-Hubbard transition
is indeed continuous at $U_{{\rm c},1}\equiv U_{{\rm c}}$
without a discontinuity in the average double occupancy.

Within perturbation theory,
the ground-state energies from weak and strong coupling do not become equal.
Further terms in the $1/U$ expansion are also negative, 
so that eq.~(\ref{E01overU}) appears to give an upper bound
to the ground-state energy. On the other hand, 
the perturbation expansion in~$U$ gives a maximum
in $E_0(U)$ at $U=1.12[1] W$ and thus provides a lower
bound around $U=W$. Therefore, comparing energies
from perturbation expansions will always result
in a lower ground-state energy for the `metal'
(expansion in~$U$) than for the `insulator'
(expansion in $1/U$). In the above case we find
the minimum energy difference
at $U_{\rm cross}$, $\Delta E_0(U_{\rm cross})/L=0.018[1] W$.

Further observables will be discussed in Sects.~\ref{sec:DMFT}
and~\ref{sec:RDA}.

\section{Dynamical Mean-Field Theory (DMFT)}
\label{sec:DMFT}

In this section, we summarize
the Dynamical Mean-Field Theory and
two of its numerical implementations. The self-consistent solution
of the single-impurity Anderson model is carried out by
the Fixed-Energy Exact Diagonalization 
(FE-ED)~\cite{EastwoodGebhard} and the
Dynamical Density-Matrix Renormalization Group 
(DDMRG)~\cite{Jeckelmann,JeckelNishimoto}.
We compare our results for the density of states
from perturbation theory with those from FE-ED, DDMRG, 
the Numerical Renormalization Group (NRG)~\cite{BullaPRL},
and Iterated Perturbation Theory (IPT).

\subsection{Single-Impurity Anderson Model}
\label{subsec:SIAM}

In the limit of infinite dimensions~\cite{MV} and under the conditions
of translational invariance and convergence of perturbation theory in strong
and weak coupling, lattice models for correlated electrons 
can be mapped onto single-impurity models~\cite{BrandtMielsch,Jarrell,RMP},
which must then be solved self-consistently. 
In general, these impurity models cannot be solved analytically.

For an approximate numerical treatment various different implementations
are conceivable.
One realization is the single-impurity Anderson model in the `star geometry',
\begin{eqnarray}
\hat{H}_{\rm SIAM} &=&\sum_{\ell=1;\sigma}^{n_s-1}
\epsilon_{\ell} \hat{\psi}_{\sigma;\ell}^+\hat{\psi}_{\sigma;\ell}
+ U \left( \hat{d}_{\uparrow}^+\hat{d}_{\uparrow} -\frac{1}{2}\right)
\left( \hat{d}_{\downarrow}^+\hat{d}_{\downarrow} -\frac{1}{2}\right) 
\nonumber \\
&& + \sum_{\sigma} \sum_{\ell=1}^{n_s-1} V_{\ell}
\left( \hat{\psi}_{\sigma;\ell}^+\hat{d}_{\sigma} + 
\hat{d}_{\sigma}^+\hat{\psi}_{\sigma;\ell}\right) \; ,
\label{SIAMns}
\end{eqnarray}
where $V_{\ell}$ are real, positive hybridization matrix elements.
The model describes the hybridization of an impurity site 
with an on-site Hubbard interaction to $n_s-1$ bath sites without interaction 
at energies $\epsilon_1 <\epsilon_2 <\ldots < \epsilon_{n_s-1}$.
In order to preserve particle-hole symmetry, we must
choose $\epsilon_{n_s-\ell}=-\epsilon_{\ell}$, and
$V_{n_s-\ell}=V_{\ell}$ for $\ell=1,\ldots,n_s-1$.
For even $n_s$, this implies that there is
an energy level at $\epsilon_{n_s/2}=0$,
i.e., at the impurity level. For this bath site, which is
absent for odd $n_s$, we expect
particularly strong hybridization with the impurity level,
so that large odd-even effects can be expected.

For a given set of parameters $(\epsilon_{\ell}, V_{\ell})$
the model~(\ref{SIAMns}) defines a many-body problem for which the 
one-particle Green function 
\begin{equation}
G_{\sigma}^{(n_s)}(t) = -\I \left\langle \hat{{\cal T}} \left[
\hat{d}_{\sigma}(t) \hat{d}_{\sigma}^+\right]
\right\rangle_{\rm SIAM} 
\label{GSIAMfinite}
\end{equation}
must be calculated numerically.
Here, $\langle \ldots\rangle_{\rm SIAM}$ 
implies the ground-state expectation value within the single-im\-purity 
Anderson model.
The various implementations
of the DMFT
differ in the choice of this `impurity solver'~\cite{Vollireview},
denoted, e.g., as DMFT(FE-ED) or DMFT(DDMRG).

Ultimately, we will be interested in the limit $n_s \to\infty$ where the 
hybridization function
\begin{equation}
{\cal H}^{(n_s)}(\omega) 
= \sum_{\ell=1}^{n_s-1} 
\frac{V_{\ell}^2}{\omega-\epsilon_{\ell}+\I \eta \sgn(\omega)} 
\end{equation}
is required to approach 
the hybridization function of the \emph{continuous} problem smoothly,
\begin{equation}
{\cal H}(\omega) = \lim_{n_s\to\infty}{\cal H}^{(n_s)}(\omega) \; .
\end{equation}
Correspondingly, the Green function is required to fulfill
\begin{equation}
G_{\sigma}(\omega)
= \lim_{n_s\to\infty}
G_{\sigma}^{(n_s)}(\omega) \; .
\end{equation}
At self-consistency the Green function of the impurity problem
describes the Hubbard model in infinite dimensions.
As shown in~\cite{RMP},
the hybridization function must obey 
the simple relation
\begin{equation}
{\cal H}(\omega) = G_{\sigma}(\omega)
\label{selfcons}
\end{equation}
on the Bethe lattice.
This equation closes the self-con\-sist\-ency cycle for the continuous problem:
bath energies and hybridizations must be chosen in such a way that the
one-particle Green function and 
the hybridization function fulfill~(\ref{selfcons}).

\subsection{Fixed-Energy Exact Diagonalization (FE-ED)}
\label{subsec:FE-ED}

In the FE-ED~\cite{EastwoodGebhard} for the metallic state, 
we choose to resolve the
density of states within the frequency interval $|\omega|\leq W^*/2$.
To this end, we partition the effective bandwidth $W^*$ into
equidistant intervals of width $\delta W=W^*/(n_s-1)$, 
\begin{equation}
I_{\ell} =[-W^*/2 + (\ell-1)\delta W, -W^*/2 + \ell\delta W]
\; , \; 1\leq \ell\leq n_s-1\; .
\end{equation}
The energy levels are now fixed at the centers of $I_{\ell}$,
\begin{equation}
\epsilon_{\ell} = -W^*/2 + (\ell-1/2) \delta W \; , \; 
1\leq \ell\leq n_s-1\; .
\end{equation}
For our calculations we choose $W^*=9t=2.25 W$ which is sufficient to
resolve the main features of the density of states (quasi-particle
peak, Hubbard bands) for weak to intermediate coupling strengths,
$U\leq W$. In the FE-ED, we can be sure that our resolution increases
systematically as a function of $1/n_s$ which cannot be guaranteed
in the Caffarel-Krauth implementation of the
exact diagonalization~\cite{RMP,Krauth,BullaLDMFT,CK}.

For $n_s \leq 15$, we determine the impurity Green function
using the (dynamical) Lanczos technique.
The imaginary part of the Green function displays
$n_{\rm L}$ peaks depending on the number $n_{\rm L}\approx 100$ states 
kept in the Lanczos diagonalization. We apply a constant broadening 
of width $\delta W$ to the individual peaks 
in $\Im G_{\sigma}^{(n_s)}(\omega_r)$, 
$r=1,\ldots,n_{\rm L}$.
We then collect the weight 
into the intervals $I_{\ell}$ and assign this weight
to $w_{\ell}=V_{\ell}^2$. 
Typically, the weight of the peaks at energies
$|\omega_r|> W^*/2$ is very small. Thus we set 
\begin{eqnarray}
w_{\ell}
&=& \int_{I_{\ell}} \diff \omega 
\sum_{r=1}^{n_{\rm L}} 
\Im G_{\sigma}^{(n_s)}(\omega_r) \nonumber
\\
&& \frac{\Theta(\omega - \omega_r+ \delta W/2)
- \Theta(\omega - \omega_r- \delta W/2)}{\delta W}
\; .
\label{recollectionofweight}
\end{eqnarray}
In order to generate an educated input guess for $V_{\ell}=\sqrt{w_{\ell}}$,
we use the results from fourth-order perturbation theory
in~(\ref{recollectionofweight}).
In the FE-ED, the self-consistency cycle is stable, i.e.,
different input choices result in the same self-consistent solution
for a given $n_s$. This is not guaranteed in the Caffarel-Krauth
exact-diagonalization scheme, as shown for the Mott-Hubbard insulator
in~\cite{EastwoodGebhard}.
The reason for the stability of the FE-ED is 
that the energies $\epsilon_{\ell}$ are kept fixed.
The $n_s$ substantial peaks in 
$\Im G_{\sigma}^{(n_s)}(\omega_r)$ 
carry sufficient information to determine the $n_s$ hybridization
matrix elements $V_{\ell}$. However, these $n_s$ peaks are not sufficient
to determine $2n_s$ parameters $(\epsilon_{\ell},V_{\ell})$ uniquely.

\begin{figure}[htb]
\resizebox{8cm}{!}{\includegraphics{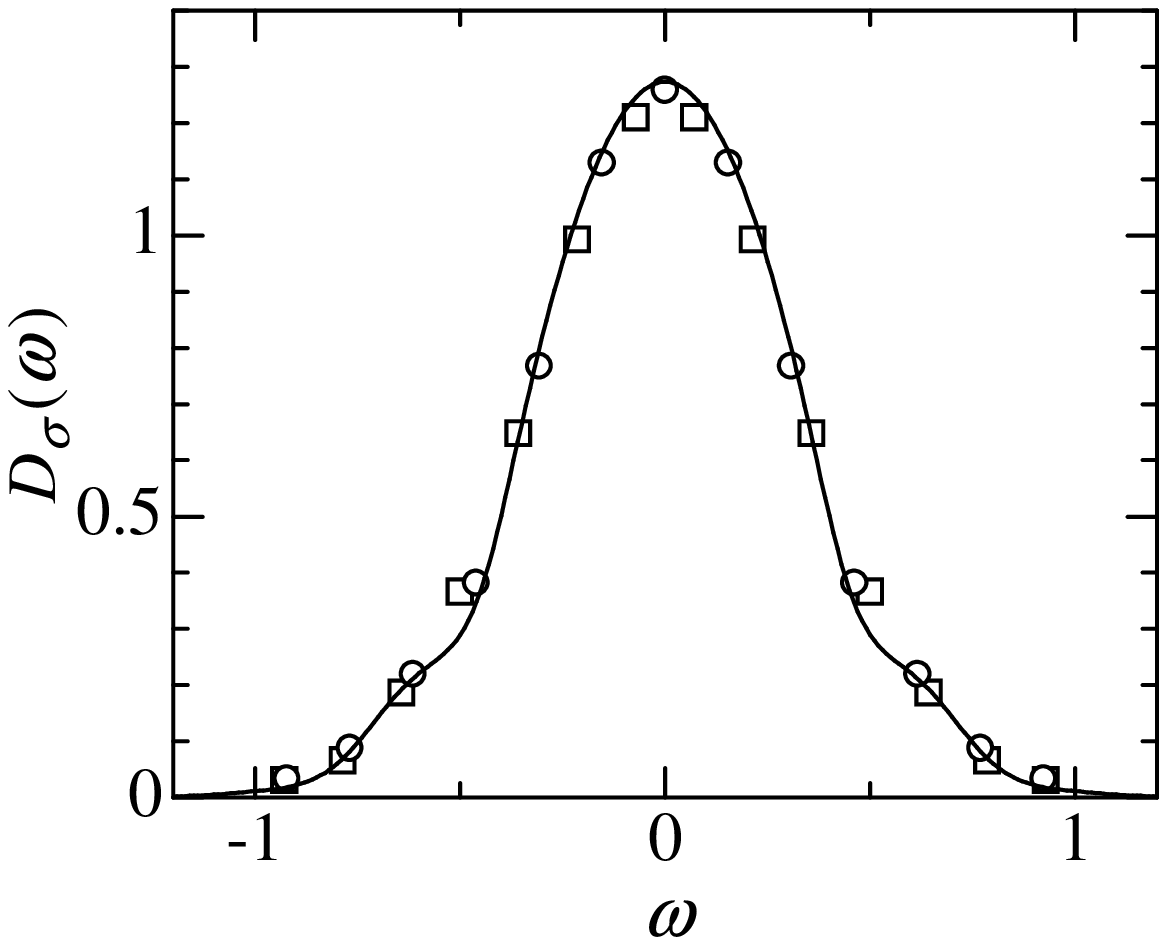}}
\vspace{3pt}
  \caption{Hybridization function ${\cal H}^{(n_s)}(\omega)$
for $n_s=14$ (circles) and $n_s=15$ (squares) in comparison
with the density of states from fourth-order perturbation theory
at $U=0.4 W$ (solid line); units $W=4t \equiv 1$.\label{Fig:FEED14_15U0.4}}
\end{figure}

Fig.~\ref{Fig:FEED14_15U0.4} shows the converged hybridization function
in FE-ED for $n_s=14$ and $n_s=15$ for 
$U=0.4 W=1.6 t$, in comparison with the results
for the density of states from fourth-order perturbation theory.
It is seen that the agreement is very good.

\begin{figure}[htb]
\resizebox{8cm}{!}{\includegraphics{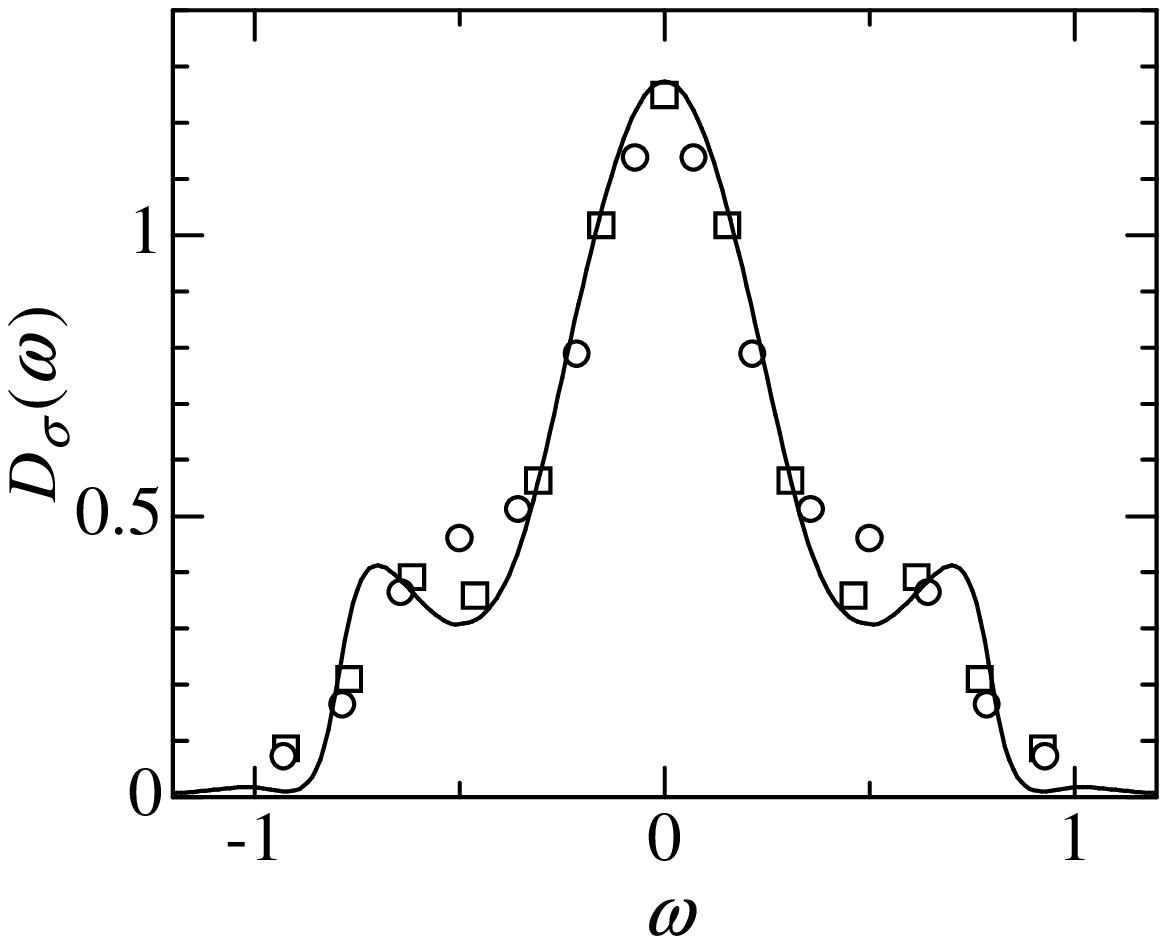}}
\vspace{3pt}
\caption{Hybridization function ${\cal H}^{(n_s)}(\omega)$
for $n_s=14$ (circles) and $n_s=15$ (squares) in comparison
with the density of states from fourth-order perturbation theory
at $U=0.6 W$ (solid line); units $W=4t \equiv 1$.\label{Fig:FEED14_15U0.6}}
\end{figure}

As seen
in Fig.~\ref{Fig:FEED14_15U0.6}, deviations 
are noticeable for $U=0.6W=2.4 t$.
Unfortunately, the resolution $\delta W=9t/14\approx 0.16 W$ 
is still rather limited, and finite-size effects are appreciable
around $\omega=0.5 W$. 
Therefore, it is unclear whether the differences
in $D_{\sigma}(\omega)$ between FE-ED and perturbation theory
around $\omega=0.5 W$ are due to finite-size effects of the FE-ED, or
due to missing sixth-order terms in the perturbation expansion. 
The DDMRG which is able
to treat much bigger systems allows us to resolve this issue.

\subsection{Dynamical Density-Matrix Renormalization Group (DDMRG)}
\label{subsec:DDMRG}

The Density-Matrix Renormalization Group (DMRG)
is a numerical method in which the energy functional 
\begin{equation}
E(|\Psi\rangle) = \langle \Psi | \hat{H} | \Psi \rangle
\end{equation}
is minimized variationally in the subspace
of normalized states
$\langle \Psi | \Psi \rangle=1$. 
This is done by carrying out a numerical diagonalization on a system
of finite size and a renormalization on a part of this system.
At each step of the renormalization, $m$ density-matrix eigenstates of
the subsystem
are kept to build a variational basis of dimension ${\cal O}(m^2)$;
for a review, see~\cite{DMRGbook}.
The optimal state for a given $m$, $|\Psi\rangle_{\rm opt}$, provides
a variational bound for the ground-state energy,
$E_0^{\rm var} ={}_{\rm opt} \langle \Psi | \hat{H} | \Psi \rangle_{\rm opt}$.
For one-dimensional lattice systems, 
this method provides a highly accurate estimate for
ground-state properties of hundreds of interacting electrons.
The accuracy of the energy, $E_0^{\rm var} = 
E_0 + {\cal O}(\varepsilon^2)$, is better than that of the
variational ground state itself,
$|\Psi\rangle_{\rm opt} =|\Psi_0\rangle + {\cal O}(\varepsilon)$.
Typically, the error $\varepsilon^2$ scales as the weight
$P_m$ of the discarded density-matrix eigenstates.

In DDMRG~\cite{Jeckelmann}, this concept is generalized to
the minimization of a frequency-dependent functional.
For the local Green function and $\omega\geq 0$, it reads
\begin{eqnarray}
W_{\eta}(| \Psi\rangle ;\omega) &=& 
\langle \Psi | \left(E_0+\omega-\hat{H}\right)^2 +\eta^2| \Psi \rangle
\nonumber \\
&& + \eta \langle \Psi_0 | \hat{c}_{\sitei,\sigma} | \Psi\rangle
+ \eta \langle \Psi | \hat{c}_{\sitei,\sigma}^+ | \Psi_0\rangle
 \; .
\end{eqnarray}
Here the frequency $\omega$ is fixed for an individual DDMRG run.
The optimal state $|\Psi\rangle_{\rm opt}$ is the imaginary part
of the so-called correction vector. The optimal functional becomes
\begin{eqnarray}
W_{\eta}^{\rm opt}(\omega) &=& -\eta^2 \langle \Psi_0 |
\hat{c}_{\sitei,\sigma} 
\Bigl[  
\left(E_0+\omega-\hat{H}\right)^2 +\eta^2
\Bigr]^{-1} 
\hat{c}_{\sitei,\sigma}^+
|\Psi_0\rangle 
\nonumber \\
&=& -\eta ^2 \sum_{n} 
\frac{\left|\langle \Psi_0 | \hat{c}_{\sitei,\sigma} |\Psi_n\rangle\right|^2}%
{\left(E_0+\omega-E_n\right)^2 +\eta^2}
\; ,
\end{eqnarray}
where $|\Psi_n\rangle$ and $E_n$ are the exact eigenstates and energies
of the Hamiltonian. The spectral representation~(\ref{spectralrepresentation}) 
shows that, in the thermodynamic limit and up to a factor of $-\pi\eta$, 
the DDMRG provides the exact local density of states at frequency $\omega$,
convolved with a Lorentzian of width $\eta$,
\begin{equation}
D_{\sigma}^{\eta}(\omega) = -\frac{1}{\pi \eta} W_{\eta}^{\rm opt}(\omega) \; .
\end{equation}
The main advantage of this variational approach is that, as in the 
ground-state energy calculations, the optimal value of
the functional (i.e., the density of states) is obtained with an accuracy
of the order of $\varepsilon^2$ if the optimal state $|\Psi\rangle_{\rm opt}$
is calculated with an accuracy $\varepsilon$. 

As the DMRG method is most accurate for systems with a quasi one-dimensional 
structure,
we perform calculations of the single-impurity Anderson model~(\ref{SIAMns})
in its equivalent linear-chain form~\cite{Costi}
\begin{eqnarray}
\hat{H}_{\rm SIAM} &=&
 U \left( \hat{d}_{\uparrow}^+\hat{d}_{\uparrow} -\frac{1}{2}\right)
\left( \hat{d}_{\downarrow}^+\hat{d}_{\downarrow} -\frac{1}{2}\right) 
\nonumber \\
&& + V \sum_{\sigma} \left( \hat{f}_{\sigma;0}^+\hat{d}_{\sigma} +
\hat{d}_{\sigma}^+\hat{f}_{\sigma;0}\right) 
\label{SIAMchain}  \\
&& + \sum_{\sigma} \sum_{\ell=0}^{n_s-2} \lambda_{\ell}
\left( \hat{f}_{\sigma;\ell}^+\hat{f}_{\sigma;\ell+1} + 
\hat{f}_{\sigma;\ell+1}^+\hat{f}_{\sigma;\ell}\right) \; . \nonumber
\end{eqnarray}
The DDMRG provides the exact local density of states for a finite chain
with $n_s$ sites. To obtain the spectrum of an infinite chain,
the broadening $\eta$ must be scaled as a function
of the system size~\cite{Jeckelmann}. 
If $\eta$ is chosen too small, the DDMRG density of states displays 
finite-size peaks. If $\eta$ is chosen too large, relevant
information is smeared out.
As an empirical fact, 
$\eta\approx W^*/n_s$ should be chosen, i.e., the resolution
scales as the inverse system size, as found for one-dimensional
lattice models. 
The DDMRG technique for the single-impurity Anderson model
will be described in more detail elsewhere~\cite{JeckelNishimoto}.

In the DDMRG extension of the FE-ED, the limitations of the
Lanczos technique are overcome in two ways. First, we 
may use bigger systems. The DDMRG can handle $n_s=64$ sites
on a workstation ($m=300$ states kept in the renormalization)
with CPU time as the limiting factor
(48~hrs on a 500~MHz DEC-alpha workstation). In contrast, the exact
diagonalization studies are seriously limited by memory constraints
(3~GByte of memory for $n_s=15$).

Second, the DDMRG provides the density of states at selected
frequencies $\omega_i$. Typically, we choose them to resolve the effective
bandwidth~$W^*$ equidistantly, $\omega_{i+1}-\omega_i= \delta \omega 
\approx \eta\approx \delta W$.
We then `deconvolve' the DDMRG data by inverting the Lorentz transformation
\begin{equation}
D_{\sigma}^{\eta}(\omega_i) = \sum_{j} \frac{\delta \omega}{\pi}
\frac{\eta}{\eta^2 + (\omega_i-\omega_j)^2} 
D_{\sigma}(\omega_j) \; .
\end{equation}
The procedure can be repeated for different choices of the equidistant
frequencies~$\omega_i$ to get more values of $D_{\sigma}(\omega_j)$.
In this way, the DDMRG provides a set of values $D_{\sigma}(\omega_j)$
for the density of states 
which, for the same $n_s$, is at least as dense as the set from
the Lanczos diagonalization. In practice, we use from one to four
different sets of frequencies, corresponding to a frequency resolution
from $\eta$ to $\eta/4$. Naturally, structures with an intrinsic width 
of less than $\eta$ cannot be resolved with this procedure even if
we use different sets of frequencies. 
The main advantage of this transformation, however, 
is that no extrapolation or scaling
analysis of these values $D_{\sigma}(\omega_j)$ is necessary because they
converge very quickly to the $n_s \to \infty$ limit.
Therefore, with DDMRG we obtain a more accurate discrete representation
of the density of states for a given $n_s$ than within the
Lanczos FE-ED.

\begin{figure}[htb]
\resizebox{8cm}{!}{\includegraphics{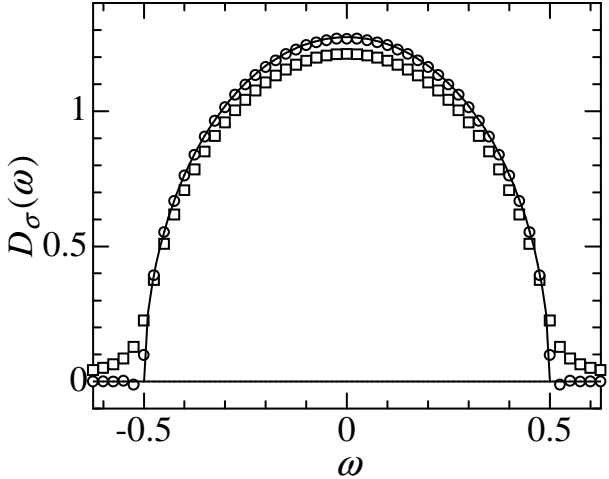}}
  \caption{Density of states for $U=0$ from DDMRG on $n_s=64$ sites, 
compared with the bare density of states
$\rho_0(\omega)$~(\protect\ref{rhozero}) (full line); units $W=4t \equiv 1$.
Squares: data from DDMRG for $\delta\omega=\eta=0.1t$.
Circles: DDMRG result after deconvolution.\label{Fig:DDMRGzero}}
\end{figure}

An example for the non-interacting case is shown in
figure~\ref{Fig:DDMRGzero}, where we have chosen 
$n_s=64$, $W^*=5t$, $\delta W=W^*/63$,
and $\delta\omega=\eta=0.1t$.
This is a relevant test of accuracy because the non-interacting
single-impurity Anderson model poses a non-trivial problem to
DDMRG~\cite{JeckelNishimoto}.
As seen in figure~\ref{Fig:DDMRGzero},
there is an excellent agreement between the 
`deconvolved' numerical data for $n_s=64$ 
and the exact result for $n_s \rightarrow \infty$.

\begin{figure}[htb]
\resizebox{8cm}{!}{\includegraphics{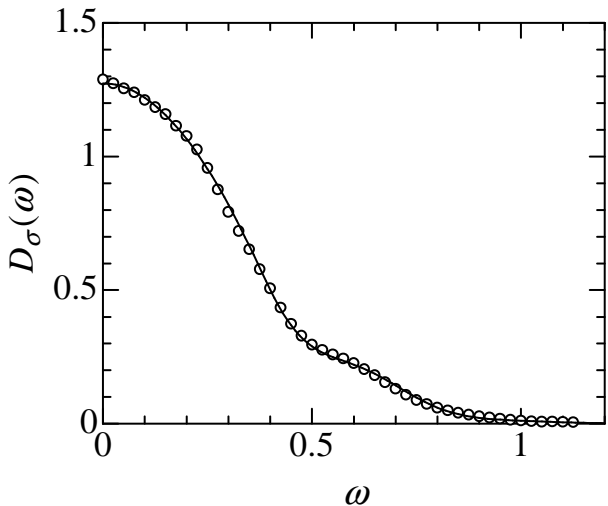}}
  \caption{Density of states for $U=0.4W$ from DDMRG on $n_s=32$ sites
after deconvolution (circles), 
compared with fourth-order perturbation theory
(full line); units $W=4t \equiv 1$.\label{Fig:DDMRG04}}
\end{figure}

\begin{figure}[htb]
\resizebox{8cm}{!}{\includegraphics{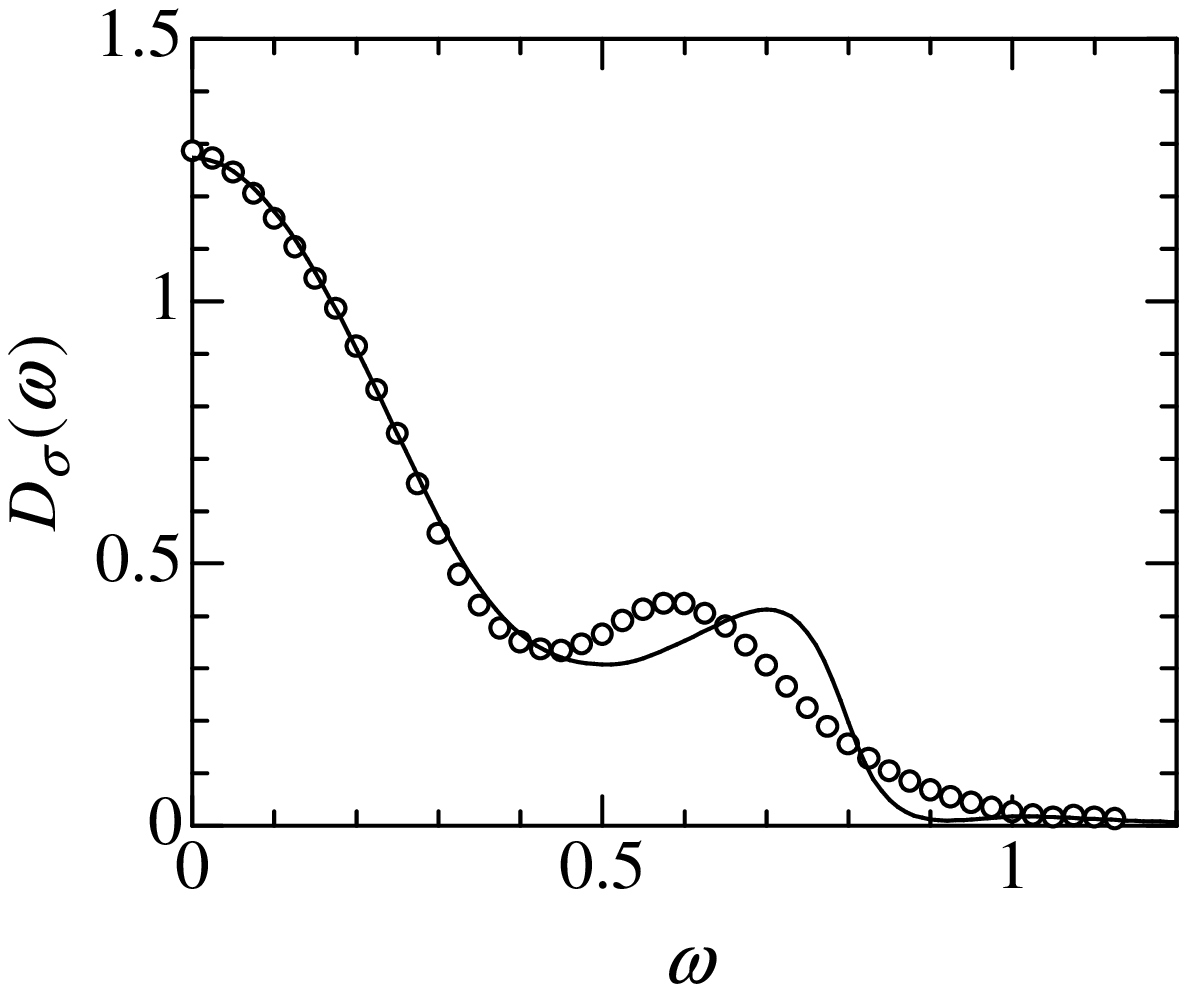}}
  \caption{Density of states for $U=0.6 W$ from DDMRG on $n_s=32$ sites
after deconvolution (circles), compared with fourth-order perturbation theory
(full line); units $W=4t \equiv 1$.\label{Fig:DDMRG06}}
\end{figure}

The density of states from DDMRG and fourth-order perturbation theory
also agree very nicely for $U=0.4W$, as is seen in 
figure~\ref{Fig:DDMRG04}. This result confirms the reliability
of both the DDMRG and our perturbation theory at weak coupling.

The agreement is not perfect for $U=0.6W$, as seen in 
figure~\ref{Fig:DDMRG06}. The DDMRG is applicable at all
interaction strengths, it is limited only by finite-size
effects (resolution $\eta\propto 1/n_s$).
Therefore, the deviations must be attributed to
the sixth-order contributions which are missing 
in our perturbation expansion. The magnitude of the discrepancies
around $\omega=0.7 W$ is indeed consistent with corrections of the 
order $0.6^2=36\% $ relative to the fourth-order term,
which dominates the second-order term around these frequencies.
Therefore, we are confident that the DDMRG provides reliable
results for all interaction strengths within its resolution limitations.

\subsection{Numerical Renormalization Group (NRG)}
\label{sec:NRG}

\begin{figure}[htb]
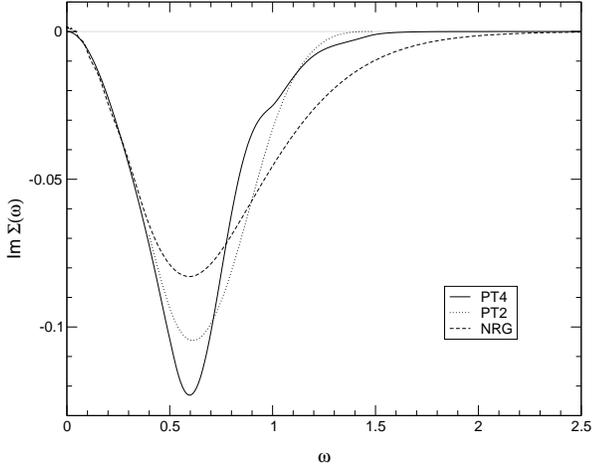

\centerline{
\resizebox{7.8cm}{!}{\includegraphics{Vgl04ImS_PT_NRG.eps}}
}

\vspace*{12pt}

\centerline{
\resizebox{7.8cm}{!}{\includegraphics{Vgl04ReS_PT_NRG.eps}}
}
\caption{Real and imaginary parts of the self-energy from NRG for $U=0.4 W$
(dashed line), compared with fourth-order perturbation theory
(full line)~\cite{Bullaprivate}. 
Also shown is the second-order self-energy (dotted line);
units $W=4t \equiv 1$.\label{Fig:NRGSE04}}
\end{figure}

The Numerical Renormalization Group (NRG) is a technique which
aims to resolve accurately the self-energy and the density of
states near $\omega=0$. To this end, an exact diagonalization
of the single-impurity model is performed at the $n$-th step
of the renormalization procedure for a frequency-interval 
$I_{\omega}^n$ around $\omega=0$. The width of the frequency
interval is cut in half
at the next step of the renormalization procedure, starting at
$I_{\omega}^0=W^*$. Therefore, features near $\omega=0$ are
resolved with exponentially increasing accuracy whereas
higher frequencies are represented by a few peaks
which are broadened on a logarithmic scale~\cite{BullaPRL}.
As a result of this logarithmic mesh for the bath energies
$\epsilon_{\ell}$, frequencies of the order of the Hubbard bands
are resolved with a rather limited accuracy.
Note that weight from the broadened Hubbard bands enters the
low-energy physics again through the iterative self-consistency 
procedure. 

Fig.~\ref{Fig:NRGSE04} shows the comparison of the self-energies from NRG and
fourth-order perturbation theory for $U=0.4 W$. 
At intermediate to large frequencies, the NRG does not reproduce 
the fingerprints of the Hubbard bands as already seen
in perturbation theory.
The same quantitative differences are found for the real part of
the self-energy. 

As a consequence, the density of states
in NRG does not show any signs of the Hubbard bands at $U=0.4 W$,
in contrast to the exact result. 
Fig.~\ref{Fig:NRGDOS} shows the density of states from NRG as
compared with perturbation theory for $U=0.4 W$ and $U=0.6 W$.
Even at $U=0.6 W$, where the Hubbard bands are clearly visible,
the NRG displays only a small shoulder. We conclude that the NRG
has problems at frequencies of the order of the Hubbard bands
whose height and positions cannot be determined reliably.

\subsection{Iterated Perturbation Theory (IPT)}
\label{sec:IPT}

\begin{figure}[htb]
\centerline{
\resizebox{7.8cm}{!}{\includegraphics{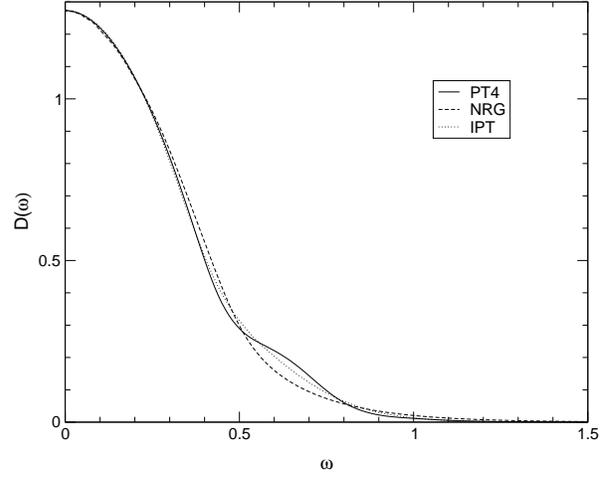}}
}

\vspace*{12pt}

\centerline{
\resizebox{7.8cm}{!}{\includegraphics{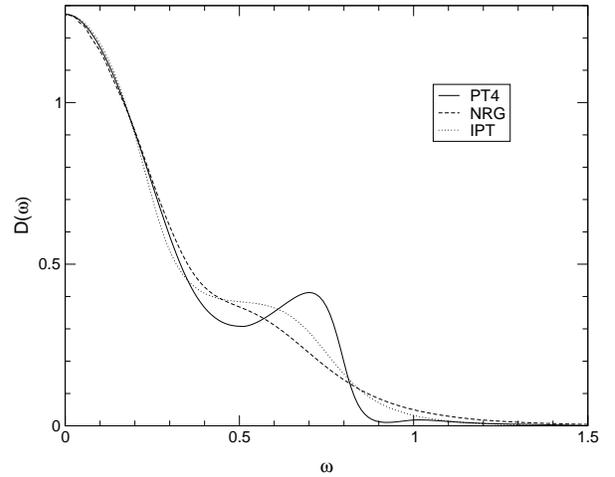}}
}
  \caption{Density of states for $U=0.4 W$ and $U=0.6 W$
from NRG (dashed lines)~\cite{Bullaprivate}, 
IPT (dotted lines)~\cite{Eastwoodprivate}, and 
fourth-order perturbation theory (full lines); units $W=4t \equiv 1$.
\label{Fig:NRGDOS}\label{Fig:IPTDOS}}
\end{figure}

The IPT approximation to
the self-energy of the 
$Z\rightarrow\infty$ Bethe lattice is given by~\cite{KotliarIPT,Mikethesis}
\begin{equation}
\Sigma_{\sigma}^{\rm IPT}(\omega)=
U^2\int_{-\infty}^{\infty}\frac{\diff\Omega}{2\pi \I} 
\Pi(\Omega){\cal G}_{\sigma}(\omega-\Omega)\; .
\label{sipt}
\end{equation}
Here, ${\cal G}_{\sigma}(\omega)$ is the host Green function 
\begin{equation}
\label{scrgipt}
{\cal G}_{\sigma}(\omega)=\frac{1}{\omega-G_{\sigma}(\omega)}
\end{equation}
and we have defined
\begin{equation}
\label{piipt}
\Pi(\omega)= - \int \frac{\diff\omega_1}{2\pi \I}
{\cal G}_{\sigma}(\omega_1)
{\cal G}_{\sigma}(\omega_1-\omega)
\end{equation}
as the polarization bubble of the host Green functions.

By construction, IPT reduces to second-order perturbation theory
in the limit of weak coupling. However, it omits the fourth-order
diagrams with vertex corrections shown in figures~\ref{Fig:4thorderdiagramB}
and~\ref{Fig:4thorderdiagramC}, whereas it weighs the other
diagrams differently. Thus, it is uncontrolled at moderate
interaction strengths.

Fig.~\ref{Fig:IPTDOS} shows the density of states from IPT for
$U=0.4 W$ and $U=0.6 W$ in comparison with fourth-order
perturbation theory. It is seen that the IPT does surprisingly well
at weak coupling considering that it is exact only to ${\cal O}(U^2)$.
 In this limit and for half band-filling,
it appears to be superior 
to the results from NRG. However, it does not reproduce the 
height of the Hubbard bands correctly.

\section{Random Dispersion Approximation}
\label{sec:RDA}

In this section,
we present numerical results from the 
Random Dispersion Approximation (RDA). 
It becomes exact for lattice electrons in high dimensions.
In contrast to the FE-ED, DDMRG, or NRG, it is not based on the DMFT 
self-consistency equations. Moreover, the RDA does not require
the convergence of the perturbation expansion.
Therefore, it provides an independent check of the validity
of perturbation theory and the DMFT approach.

\subsection{Method}

In the Random Dispersion Approximation, the dispersion relation
$\epsilon(\sitek)$
in the kinetic energy is replaced by a random quantity 
$\epsilon^{\rm RDA}(\sitek)$
where the bare density of states acts as the probability distribution,
\begin{equation}
\rho(\epsilon) = \frac{1}{L} 
\sum_{\sitek} \delta(\epsilon-\epsilon^{\rm RDA}(\sitek))
\; ,
\label{dosdefRDA}
\end{equation}
and all correlation functions factorize according to~(\ref{RDADq}), etc.
This is the characteristic property of the dispersion relation in
infinite dimensions~\cite{Gebhardbook,RDA}, so that the RDA 
with the semi-elliptic density of states~(\ref{rhozero})
becomes
exact for the Bethe lattice with infinite coordination number.

In order to put this idea into practice, we choose a one-dimensional 
lattice of $L$ sites in momentum space 
\begin{equation}
k_{\ell} = \frac{2\pi}{L} \left( -\frac{L+1}{2} +\ell \right) \;  , \; 
(\ell = 1,\ldots,L) 
\; ,
\end{equation}
and determine the dispersion relation
$\epsilon (k)$ as the solution of the implicit equation
\begin{equation}
k/2 = \left(2\epsilon(k)/W\right)
[1-\left(2\epsilon(k)/W\right)^2]^{1/2}
+\arcsin\left(2\epsilon(k)/W\right)
\; .
\end{equation}
This choice guarantees $\rho(\epsilon)=\rho_0(\epsilon)$
in the thermodynamic limit.

Next, we choose a permutation ${\cal Q}_{\sigma}$ for each
spin direction~$\sigma$ which permutes
the sequence $\{1,\ldots,L\}$ 
into $\{{\cal Q}_{\sigma}[1],\ldots,{\cal Q}_{\sigma}[L]\}$.
This defines a realization of the RDA dispersion,
${\cal Q}=[{\cal Q}_{\uparrow},{\cal Q}_{\downarrow}]$.
The numerical task is then the Lanczos diagonalization of
the Hamiltonian
\begin{equation}
\hat{H}^{{\cal Q}}
= \sum_{\sigma}\sum_{\ell=1}^{L} \epsilon(k_{{\cal Q}_{\sigma}[\ell]})
\hat{c}_{k_{\ell},\sigma}^+\hat{c}_{k_{\ell},\sigma}
 + U \hat{D}
\; .
\end{equation}
In this way we obtain the momentum distribution
\begin{equation}
n^{\cal Q}(\epsilon;U)=\frac{1}{2} \sum_{\sigma}
\left. \vphantom{\sum}
\langle \hat{n}_{k_{\ell},\sigma} \rangle
\right|_{\epsilon(k_{\ell})=\epsilon} \; ,
\end{equation}
where $\langle \ldots \rangle$ denotes the ground-state 
expectation value for the realization~${\cal Q}$.

As a next step, we obtain all physical quantities
for fixed system size~$L$ by averaging over
$N_{\cal Q}$ realizations~${\cal Q}$.
Typically, we choose at least $N_{\cal Q}=100$ for
$6 \leq L\leq 14$, and $N_{\cal Q}=50$ for $L=16$.
For the physical quantities,
we obtain gaussian-shaped distributions for which we
can determine the average values, e.g.,
\begin{equation}
n(\epsilon;U)=\frac{1}{N_{\cal Q}}\sum_{{\cal Q}} 
n^{\cal Q}(\epsilon;U)
\end{equation}
with accuracy ${\cal O}(1/N_{\cal Q})$.

In order to improve the quality of our distributions slightly,
we impose a filter on our randomly chosen permutations. 
For a truly random dispersion, $L|t^{\rm RDA}(\ell)|^2 
=\overline{\epsilon^2}$ is independent
of~$\ell$~\cite{Gebhardbook}. Therefore, we discard those
realizations for which 
\begin{equation}
\sum_{\ell=1}^{L-1} \left[|t^{{\cal Q}_{\sigma}}(\ell)|^2
- \overline{\epsilon^2} \right]^2 > d_L
\end{equation}
with $d_L \approx 0.2$ for $L\leq 16$.
In this way, we admit about every second
of the randomly chosen configurations.
Note that there are of the order of $(L!)^2$ 
different realizations so that our filter does not 
introduce any unwanted bias.

\subsection{Momentum distribution}

\begin{figure}[htb]
\resizebox{\columnwidth}{!}{\includegraphics{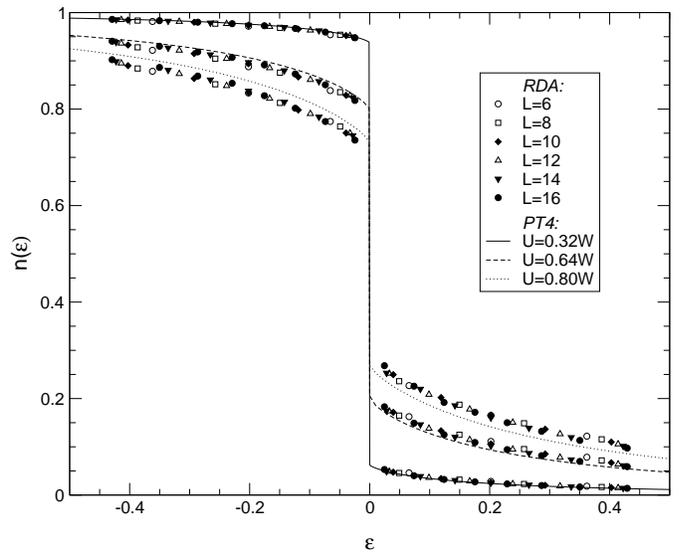}}
\caption{Momentum distribution in RDA 
for $U=W/\pi\approx 0.32 W$, $U=2W/\pi \approx 0.64 W$, 
and $U=5W/(2\pi)\approx 0.80 W$ for even system sizes $L\leq 16$.
The lines are the result from fourth-order perturbation theory.
Note the weak $\epsilon \ln\epsilon$ behavior near the 
discontinuity.\label{Fig:rdamomentum}}
\end{figure}

Fig.~\ref{Fig:rdamomentum} shows the RDA momentum distribution
in comparison with results from perturbation theory.
As seen from the figure,
finite-size effects are rather small in the metallic phase.
All values for the momentum distribution appear to fall onto
almost the same curve. For small interaction strengths,
the RDA data lie essentially on top of the perturbative results.
Deviations for larger interactions can be attributed to 
missing higher-order corrections to our fourth-order result.
Nevertheless, it is quite astonishing that fourth-order
perturbation theory provides a sensible description
of the momentum distribution for interaction strengths as
large as $U\approx 0.8 W$.  

\begin{figure}[htb]
\resizebox{\columnwidth}{!}{\includegraphics{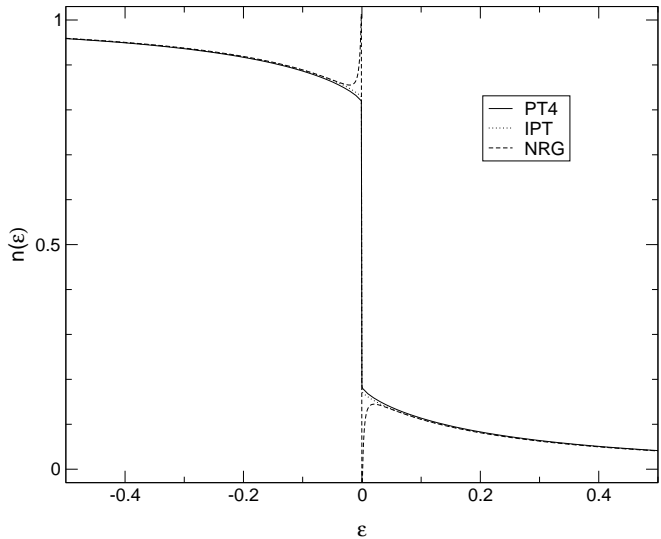}}
\caption{Momentum distribution for $U=0.6 W$ in IPT 
(dashed line)~\cite{Eastwoodprivate}, 
NRG (dotted line)~\cite{Bullaprivate}, and fourth-order perturbation theory
(full line). The deviations of NRG near the jump result 
from a small but finite positive imaginary part of the NRG self-energy 
around $\omega=0$.\label{Fig:IPTNRGmomentum}}
\end{figure}

In figure~\ref{Fig:IPTNRGmomentum} we compare our results 
for the momentum distribution at $U=0.6 W$ from
fourth-order perturbation theory 
with those from IPT and NRG. We see that both approaches
describe the momentum distribution very well
despite their shortcomings for the self-energy at intermediate
frequencies. Apparently, for small to intermediate coupling strengths, 
the overall momentum distribution is not a very sensitive test
for the quality of an approximation to the self-energy at intermediate 
frequencies. 

\begin{figure}[htb]
\resizebox{\columnwidth}{!}{%
\includegraphics{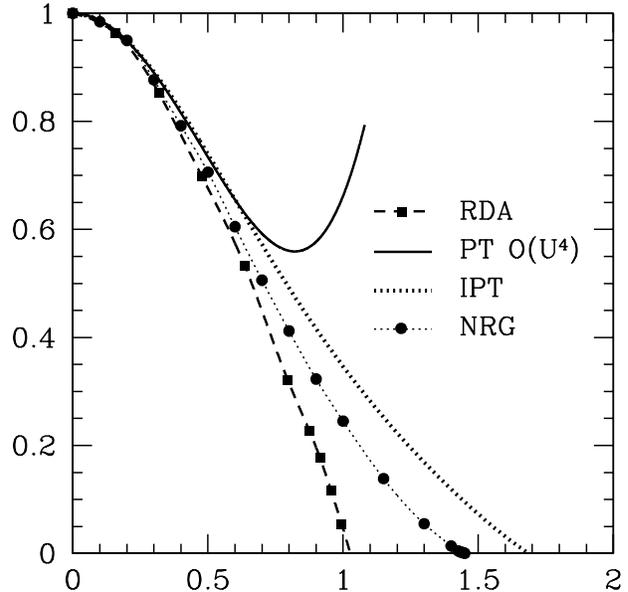}}
\caption{Quasi-particle weight as a function of $U/W$
in IPT (dotted line)~\cite{Eastwoodprivate}, 
NRG (filled circles)~\cite{Bullaprivate}, RDA (squares),
and fourth-order perturbation theory (full line).\label{Fig:Zfactor}}
\end{figure}

A more sensitive quantity is the discontinuity of the momentum
distribution at $\epsilon=0$, see~(\ref{nepsjump}).
In figure~\ref{Fig:Zfactor} we show the quasi-particle weight
as a function of~$U/W$ as obtained from IPT, NRG, RDA, and
fourth-order perturbation theory. For the latter quantity, we invert
eq.~(\ref{4thZ}) which gives
\begin{equation}
Z(U) = 1 - 1.307[1] \left(\frac{U}{W}\right)^2 
+0.969[2] \left(\frac{U}{W}\right)^4 + {\cal O}(U^6) \; .
\label{Zfaceq}
\end{equation}
In the region where fourth-order perturbation is reliable, 
$U\leq 0.6 W$, the results agree with those from NRG and RDA 
within their numerical accuracy. Therefore, fourth-order
perturbation theory cannot discriminate in favor
of either of the two approaches which, however, support
different scenarios for the Mott-Hubbard metal-insulator
transition. 

The quasi-particle weight from Iterated Perturbation Theory 
closely follows the fourth-order
result to a point where~(\ref{Zfaceq}) definitely overestimates $Z(U)$,
e.g., for $U=0.7 W$. From this we conclude that IPT
generally overestimates the stability of the metal.

\section{Conclusions}
\label{sec:conclusions}

In this work we have calculated the one-particle
self-en\-ergy for the half-filled Hubbard model 
in the limit of infinite dimensions to fourth order
in the interaction strength. We have reduced 
the four non-equivalent 
fourth-order diagrams to sets of two-dimensional integrals over 
real, tabulated functions over finite intervals. From the
self-energy, we have derived the density of states, the ground-state energy,
the mean double occupancy, the momentum distribution, and the quasi-particle
weight, and have compared it to various analytical and numerical approaches
to the Hubbard model in infinite dimensions for the case of
a Bethe lattice with infinite coordination number, i.e.,
a semi-circular bare density of states.

Our results show that, at half filling, it is not permitted to select 
only subclasses of diagrams because all diagrams contribute equally.
The imaginary part of the self-energy to fourth order becomes positive
at some frequency for $U=0.64 W$, which limits the applicability
of our fourth-order results to moderate interaction strengths.
For the momentum distribution and, in particular, the ground-state energy 
and the mean double occupancy, the perturbation expansion appears
to be better behaved, i.e., the results remain sensible up to $U\approx W$.
A similar observation had been made for the single-impurity Anderson
model by Yamada and Yosida.

The Dynamical Mean-Field Theory (DMFT), which becomes
exact in infinite dimensions, requires the self-con\-sistent
solution of a single-impurity Anderson model.
Our results indicate that the Dynamical Density-Ma\-trix Renormalization Group 
(DDMRG) is an excellent `impurity solver'.
Where Exact Diagonalization is hampered by finite-size effects
($n_s\leq 15$), the DDMRG treats much bigger systems, up to $n_s =64$, 
and a new scheme for the deconvolution of the data 
further improves the frequency resolution.
In this way, we obtain a very good agreement between the results from
DDMRG and fourth-order perturbation theory where the latter is applicable. 

The agreement is found to be less satisfactory with the results
from Numerical Renormalization Group (NRG) at intermediate frequencies.
The Hubbard bands which become discernible around $U=0.4 W$
are not resolved very well within the NRG scheme. 
It would be interesting to see whether a deconvolution of the NRG data 
could improve the agreement with fourth-order perturbation theory
at the energy scale of the Hubbard bands.
Iterated Perturbation Theory (IPT) works surprisingly well for $U\leq 0.6$
for all quantities tested. However, for larger interaction strengths,
it seriously overestimates the quasi-particle weight 
and thus the stability of the metallic state. 

The Random Dispersion Approximation (RDA), which also becomes exact
in infinite dimensions, provides the momentum distribution and
the quasi-particle weight. The results agree with those from 
fourth-order perturbation theory within the limits of its applicability. 
Unfortunately, for $U\leq W$, the momentum distribution
and the quasi-particle weight turn out not be very sensitive quantities, 
i.e., all methods equally well reproduce the results 
from fourth-order perturbation.
However, the RDA and NRG support two different scenarios for the
Mott-Hubbard me\-tal-insulator transition. 
The differences between NRG and RDA in the quasi-particle-weight
become sizable only for $U> 0.8 W$, a region which, unfortunately,
cannot be accessed with fourth-order perturbation theory.

We hope that the DDMRG approach will enable us to investigate
the region $U>0.6 W$ in more detail. Work in this
direction is in progress.

\subsubsection*{Acknowledgments}

We thank David Logan for helpful discussions.
We are grateful to Michael P.~Eastwood for his IPT
results and to Ralf Bulla for making available to us his NRG data.
We thank the HRZ Darmstadt computer facilities where some
of the calculations were performed.
Support by the Deutsche Forschungsgemeinschaft and 
the center {\sl Optodynamics\/} of the Philipps-Universit\"at Marburg 
is gratefully acknowledged.

\appendix

\section{Momentum distribution near the discontinuity}
\label{nepsapp}

Using the Luttinger relations~(\ref{FLZfactor}) and~(\ref{FLalpha})
for a Fermi liquid, we evaluate~(\ref{defnofepsilon})
for $n_{\sigma}(\epsilon>0)=(1-Z(U))/2 +\delta n(\epsilon)$.
Then, 
\begin{equation}
\delta n_{\sigma}(\epsilon) = \int_{-\omega_c}^0 \frac{d\omega}{\pi}
\frac{\gamma\omega^2}{(\omega/Z-\epsilon)^2 + \gamma^2\omega^4} + \ldots \; ,
\end{equation}
apart from regular terms which come from the incoherent background
in~(\ref{Defspectral}).
Naturally, we are interested in the behavior close to the Fermi energy,
i.e., for energies which fulfill
\begin{equation}
0 < Z(U) \epsilon < \omega_c \; .  \label{condone}
\end{equation}
Therefore, we require that
\begin{equation}
Z(U) \propto \omega_c \label{Zscaling}
\end{equation}
in order to observe a finite $\epsilon$-interval in which the formulae
below are applicable. Furthermore, we choose $\epsilon>0$
so that we are above the jump in the distribution.

As can be seen by an explicit calculation, the
dominant contribution for small $\epsilon$ comes from
the frequencies $\omega=Z(U)\epsilon$. Then, we may ignore the
term proportional to $\omega^4$ in the denominator and we find
\begin{equation}
\delta n_{\sigma}(Z(U)\epsilon\ll \omega_c) = \frac{\gamma Z(U)^2\omega_c}{\pi}
\int_{-1}^0 
\frac{\diff x\, x^2}{(x-Z(U)\epsilon/\omega_c)^2} \; .
\label{xequation}
\end{equation}
This approximation requires that
\begin{equation}
\epsilon \ll \frac{1}{Z(U)^2\gamma} \; . 
\end{equation}
This condition should correspond to~(\ref{condone}) so that
\begin{eqnarray}
\frac{1}{Z(U)\gamma} &\propto& \omega_c \; , \quad \hbox{i.e.,} \\
\gamma & \propto& (\omega_c)^{-2} \; . 
\label{alphascaling}
\end{eqnarray}
The scaling laws~(\ref{Zscaling}) and~(\ref{alphascaling}) appear to be
natural, given the Fermi liquid equations~(\ref{FLZfactor}) 
and~(\ref{FLalpha}). To leading order in $\epsilon\ln \epsilon$,
eq.~(\ref{xequation}) results in~(\ref{epslogeps}).

If the above scaling laws~(\ref{Zscaling}) and~(\ref{alphascaling})
apply, there should be a finite region around $\epsilon=0$ in which
the $\epsilon\ln \epsilon$ dependence can be seen.
However, its overall intensity scales down with an extra factor~$Z(U)$.

\section{Definition of help functions}
\label{appB}

Here we define the help functions which are necessary for
the evaluations of the diagrams to fourth order.
Explicit results are given for the semicircular density
of states~(\ref{rhozero}) in units of $W=4t \equiv 1$.

We start with two functions which characterize the
non-interacting Green function.
\begin{eqnarray}
\label{f}
f(\omega)&=&{\cal P}\int_0^{W/2}\diff\epsilon\,
\frac{\rho(\epsilon)}{\omega+\epsilon}\, , \\
\frac{\pi}{4}f(|\omega|\leq \frac{1}{2})&=&
\pi\omega-1-\sqrt{1-4\omega^2}
\ln\left|\frac{2\omega}{1+\sqrt{1-4\omega^2}}\right|\,,
\nonumber \\
&& \\
\frac{\pi}{4}f(|\omega|>\frac{1}{2})&=&
\pi\omega-1-
\sgn(\omega)\sqrt{4\omega^2-1}
\arccos\left(\frac{1}{2\omega}\right)\, ,
\nonumber \\
&& 
\end{eqnarray}
and 
\begin{eqnarray}
\label{l}
l(\omega)&=& f(\omega)-f(-\omega)  = \Re G_{\sigma}^0(\omega) \\
&=& 8\omega\left(1-\theta\left(|\omega|-1/2\right)
\sqrt{1-\frac{1}{(4\omega^2)}}\right) \; .
\nonumber \\
&& 
\end{eqnarray} 
Next, we express the imaginary part of the
bare polarization bubble in the form ($0\leq a\leq W$)
\begin{eqnarray}
\label{h}
h(a)&=&\int_0^{W/2}\diff\epsilon_1
\int_0^{W/2}\diff\epsilon_2
\, \delta(a-\epsilon_1-\epsilon_2)\rho(\epsilon_1)\rho(\epsilon_2)
\nonumber \\
&&\\
&=&\theta(\frac{W}{2}-a)\int_0^{a}\diff y
\rho\left(\frac{a+y}{2}\right)\rho\left(\frac{a-y}{2}\right)
\\
&& +\theta(a-\frac{W}{2})\int_0^{W-a}\diff y
\rho\left(\frac{a+y}{2}\right)\rho\left(\frac{a-y}{2}\right)
\nonumber  \; ,
\end{eqnarray}
and 
\begin{equation}
\Pi^0_{\sigma}(\omega)= - \int_0^{W} \diff a\, h(a)
\left(\frac{1}{\omega-a+\I\eta} - \frac{1}{\omega +a-\I\eta}\right)
\; .
\label{spectralpi}
\end{equation}
Moreover, we need the Hilbert transform of $h(a)$ 
\begin{equation}
\label{H} 
H(x)={\cal P} \int_0^{W}\diff a\frac{h(a)}{x+a}\; ,
\end{equation}
so that $\Im \Pi^0_{\sigma}(\omega\geq 0)=\pi h(\omega)$ and
$\Re \Pi^0_{\sigma}(\omega)= H(\omega)+ H(-\omega)$.

The imaginary part of the second-order self-energy requires
the function
\begin{eqnarray}
s(b)&=&\int_{0}^{W/2}\diff\epsilon_1
\int_{0}^{W/2}\diff\epsilon_2
\int_{0}^{W/2}\diff\epsilon_3 \nonumber \\
 && \rho(\epsilon_1)\rho(\epsilon_2)\rho(\epsilon_3)
\delta(b-\epsilon_1-\epsilon_2-\epsilon_3)\; ,
\label{s}
\end{eqnarray}
which leads to 
\begin{equation}
\Sigma_{\sigma}^{\rm (2)}(\omega)=
U^{2}\int_0^{3W/2}\diff b \, s(b)
\left(\frac{1}{\omega+b-\I\eta} + \frac{1}{\omega-b+\I\eta}\right)
\; ; 
\label{spectrals}
\end{equation}
compare~(\ref{ImS2result}).
Moreover, we need the Hilbert transform of $s(b)$ 
\begin{equation}
\label{Sfunc} 
S(x)={\cal P} \int_0^{3W/2}\diff b\frac{s(b)}{x+b}\; ,
\end{equation}
so that $\Im \Sigma_{\sigma}^{\rm (2)}(\omega\geq 0)=-\pi U^2  s(\omega)$ and
$\Re \Sigma_{\sigma}^{\rm (2)}(\omega)= U^2[ S(\omega)- S(-\omega)]$.

Finally, we introduce the derivative of the bare density of states
as 
\begin{eqnarray}
d(x)&=& \frac{\diff \rho(x)}{\diff x} \nonumber \\
&=& 
-\frac{16}{\pi}\frac{x}{\sqrt{1-4x^2}}\qquad\mbox{for}\quad |x|<1/2\; .
\end{eqnarray}

\end{document}